\def\year{2021}\relax
\documentclass[letterpaper]{article} 
\usepackage{aaai21}  
\usepackage{times}  
\usepackage{helvet} 
\usepackage{courier}  
\usepackage[hyphens]{url}  
\usepackage{graphicx} 
\usepackage{natbib}  
\usepackage{caption} 
\usepackage{amsmath, amsthm, amssymb, amsfonts}
\usepackage{xcolor}

\newcommand{\update}[1]{\textcolor{black}{#1}}

\usepackage[hyphens]{url}
\usepackage{booktabs}
\usepackage{algorithm}
\usepackage{algorithmic}
\usepackage{hyperref}

\usepackage{amsmath,amsfonts,bm}

\def\eqref#1{equation~\ref{#1}}

\def\1{\bm{1}}

\DeclareMathAlphabet{\mathsfit}{\encodingdefault}{\sfdefault}{m}{sl}
\SetMathAlphabet{\mathsfit}{bold}{\encodingdefault}{\sfdefault}{bx}{n}

\usepackage{subcaption}

\title{Characterizing Online Engagement with Disinformation and Conspiracies in the 2020 U.S. Presidential Election}
\author{}
\author {
    Karishma Sharma,
    Emilio Ferrara,
    Yan Liu \\
}
\affiliations {
    University of Southern California \\
    krsharma@usc.edu, 
    emiliofe@usc.edu,
    yanliu.cs@usc.edu,
}

\begin{document}

\maketitle

\begin{abstract}
Identifying and characterizing disinformation in political discourse on social media is critical to ensure the integrity of elections and democratic processes around the world. Persistent manipulation of social media has resulted in increased concerns regarding the 2020 U.S. Presidential Election, due to its potential to influence individual opinions and social dynamics. In this work, we focus on the identification of distorted facts, in the form of unreliable and conspiratorial narratives in election-related tweets, to characterize discourse manipulation prior to the election. We apply a detection model to separate factual from unreliable (or conspiratorial) claims analyzing a dataset of 242 million election-related tweets. The identified claims are used to investigate targeted topics of disinformation, and conspiracy groups, most notably the far-right QAnon conspiracy group. Further, we characterize account engagements with unreliable and conspiracy tweets, and with the QAnon conspiracy group, by political leaning and tweet types. Finally, using a regression discontinuity design, we investigate whether Twitter's actions to curb QAnon activity on the platform were effective, and how QAnon accounts adapt to Twitter's restrictions.
\end{abstract}

\section{Introduction}
\label{intro}

Disinformation and social media manipulation has threatened the integrity of elections and democracies around the world. In the 2016 U.S. Presidential Election, official investigations by the U.S. Congress revealed the presence of Russian state-backed operations attempting to manipulate the U.S. political landscape.  In the years following the investigation, social media platforms like Twitter and Facebook have actively documented several other suspicious operations linked to Russia, Iran, Venezuela, China and other countries \citep{gadde2018enabling}, suggesting presence of persistent efforts to manipulate online discussions. Although social media has been instrumental in shaping conversations around social and political issues \citep{loader2011networking}, its positive effects have been undermined by disinformation and manipulation. The promotion of disinformation, propaganda, and politically divisive narratives have been regularly seen in many social contexts \citep{sharma2020identifying,bessi2016social,martin2019trends}. 

Distortion of facts to promote disinformation and propaganda greatly reduces trust in online systems, because of its ability to influence both individual opinions and social dynamics \citep{van2015conspiracy,nyhan2010corrections}.  At an individual level, exposure to unreliable and conspiratorial narratives has been observed to influence people's perceptions of the truth and decrease pro-social behaviours \citep{van2015conspiracy,nyhan2010corrections}. Moreover, with polarized and partisan narratives, it can further exacerbate prejudices and ideological separations \citep{jolley2020exposure}. 
Secondly, the disinformation landscape is complex, with possibilities for injecting disinformation, propaganda, and influencing the discourse through automation \citep{ferrara2020characterizing}, coordination \citep{sharma2020identifying}, compromised accounts, follow trains and others \citep{torreslugo2020manufacture}. The range of activities can be widespread, with both concealed and overt efforts to promote manipulative narratives.

In the context of the 2020 U.S. 2020 presidential election, held on November 3, 2020, 
since Twitter is recognized as one of the social media platforms with the most news-focused users, with significant following in U.S. politics \citep{PewTwitter2019}, we investigate and characterize disinformation on Twitter, with an analysis of more than 240 million election-related tweets, collected between June 20, 2020 and September 6, 2020.
\update{We focus on the following research questions to characterize online engagement with disinformation and conspiracies:
\begin{itemize}
    \item \textbf{R1. What are the prevalent disinformation and conspiracy narratives on Twitter preceding the U.S. 2020 Election?} We apply and train a detection model to separate factual from unreliable (or conspiratorial) claims to examine how the political discourse was manipulated.
    \item \textbf{R2. How significant is the impact and reach of disinformation and conspiracy groups in terms of account engagements (characterized by activity level, political leaning, tweet types, and propagation dynamics)?} We characterize account engagements with the QAnon conspiracy group, and we compare propagation dynamics of unreliable/conspiracy and reliable engagement cascades.
    \item \textbf{R3. Did Twitter's restrictions on QAnon influence its activities and were they effective in limiting the conspiracy?} We investigate activity before/after the restriction and factors driving the sustained activity of QAnon accounts, and we test if QAnon accounts adapted to Twitter's restrictions using a regression discontinuity design.
\end{itemize}
}

\textbf{Contributions and findings.} The findings of the research questions studied in this paper are as follows,

\begin{itemize}
    \item We apply CSI \cite{ruchansky2017csi} to detect unreliable/conspiracy tweets in the election dataset, with validation AUC 0.8 and F1 0.76. The most prominent topics targeted by disinformation promoters preceding the election included mail-in voter fraud, COVID-19, Black Lives Matter, media censorship, and false claims and conspiracies about past and current political candidates.
    \item While disinformation tweets are widespread and diverse, engagements with such tweets are less viral than with reliable tweets (i.e., mean time to reach unique accounts is higher, mean cascade breath is smaller). Yet, there can be several such tweets which manage to receive substantial attention and engagements. 
    \item QAnon far-right conspiracies are quite prevelant on Twitter, and we find that while 85.43\% of all accounts (10.4M) in the collected data, did not have an observed engagement with QAnon accounts in the dataset, fraction of such accounts is much smaller (34.42\%) in the active 1.2 million accounts (that appear in at least 20 tweets in data).
    \item Investigation of QAnon activities suggests that their strategies include entering into active discussions with left-leaning accounts through replies, in an effort to ``red pill" (i.e., dramatically transform ones perspective with typically disturbing revelations). Also, the conspiracy group quickly adapted to Twitter's ban on its activities imposed in July, 2020. Using a regression discontinuity design to estimate causal effects of Twitter's intervention, we find statistically significant changes in hashtags adopted by QAnon. Also, the volume of engagements after the Twitter ban is still largely sustained by existing accounts created before the ban, rather than new accounts.
\end{itemize}

\update{\textbf{Research implications}. Our findings indicate that there are high interactions among more active accounts and disinformation or conspiracy groups, and engagements through reply tweets are prevalent with both left/right leaning accounts. An important implication from the study is also that the actions enforced to limit malicious groups might not be robust to evasive actions from malicious actors. Therefore, rethinking effective platform interventions is needed.} 

\update{This work is different from most previous studies, especially in investigating effectiveness of social media platform actions to limit disinformation or conspiracies. QAnon is a notable conspiracy, and most studies have traced its normification, its conspiracies and posts. Our work instead is the first to focus on characterization of interactions with QAnon accounts, and their activities and evasion of Twitter's restrictions. We also provide a large data-driven analysis of disinformation preceding the 2020 U.S. Election, focused on observed engagements for quantifying the impact of disinformation/conspiracies to better understand its risks.}


\section{Related Work}
\label{relwork}

Social media manipulation has been observed in political discourse in several countries \citep{woolley2017computational}. However, changing strategies of social media manipulation continue to challenge our understanding and findings on malicious activities on the network \citep{Cresci_2020}. Twitter reported seeing some of the operations as being less clandestine and more overt efforts to promote disinformation than they have in recent years \citep{NPRForeign}. Therefore, identification and analysis of disinformation and its landscape preceding the U.S. 2020 election is vital, along with characterization of account engagements with disinformation, to estimate its risks, and investigate the effectiveness of interventions and actions to limit it.

With regards to U.S. elections, recent papers attempt to uncover different aspects of social media manipulation.
\cite{bessi2016social} discovered tens of thousands of bot accounts distorting the 2016 election-related Twitter discourse, in what turned out to be an effort associated with the Russian Internet Research Agency \citep{badawy2019characterizing}.
\cite{luceri2020down} conducted a year long analysis of bot activity on Twitter before the 2018 midterm election. Based on detection of bot-like behaviours in  accounts, they found that 33\% of bots were active even one year prior to the midterm, and new bots were introduced into the discussion progressively up to the midterm. Nineteen thousand highly-active bots generated a large volume of 30M political messages, and humans were susceptible to retweeting bots with one-third of their retweets being content pushed by bots. \cite{ferrara2020characterizing} observed high volume of tweets from bot-like accounts prior to the November election, and found highly partisan retweeting behavior among bots and humans.  \cite{ferrara2020characterizing} also analyzed distortion with three types of conspiracy hashtags, i.e. QAnon, -gate conspiracies such as \textit{pizzagate} and \textit{obamagate}, and COVID-19 conspiracies, and found 13\% of all accounts sharing the conspiracy hashtags were suspected bots. They also found that accounts sharing content from right-leaning media were almost 12 times more likely to share conspiratorial narratives than left-leaning media accounts.

Tracking the growth of the QAnon conspiracy from fringe online subcultures to mainstream media over one year, \cite{de2020tracing} found that after the incubation period on 4chan/pol in Oct 2017, the movement had quickly migrated to larger platforms, notably YouTube and Reddit, and was covered by news media only when it moved off-line. In terms of cross-platforms effects, \cite{papasavva2020qoincidence} found spike in new account registrations on Voat after Reddit ban on QAnon threads in Sept 2018. They also studied content of QAnon posts finding discussions related to predecessor Pizzagate conspiracy, and ``Q drops", which refer to the
cryptic posts meant for adherents of the conspiracy to decode, and found more submissions but less comments in QAnon threads compared to general threads. 

Addressing growing concerns around conspiracy communities, \cite{phadke2021makes} investigated what individual and social factors promote users to join conspiracy communities. Their findings on Reddit suggest that direct interactions (i.e., replies) with conspiracists 
is the most important social precursor for joining (i.e., first contribution to) a conspiracy community. \cite{silva2020predicting}, on the other hand, investigated which factors are most predictive of engagements with factual and misleading tweets, finding follower-friend ratio, banner image or URL in user profile, presence
of image in the tweet, as most relevant, and factual tweets were more engaging in COVID-19 tweets. Differently from the papers, we focus on characterization of interactions with QAnon accounts, based on political leaning and tweet types, and on their activities and strategies to engage with other accounts and escape Twitter's restrictions.

\section{Data Collection}

For the analysis, we start from the dataset collected in \cite{chen2020election2020} which tracks election related tweets from May, 2019 onwards, and contains over approximately one billion tweets.\footnote{Dataset: \href{https://github.com/echen102/us-pres-elections-2020}{\url{https://github.com/echen102/us-pres-elections-2020}}} 
The dataset was collected by tracking mentions of official and personal accounts of Republican and Democratic candidates in the presidential election using Twitter's streaming API service, which returns matches in tweet content and metadata within a $\sim$1\% sample of the stream of all tweets. The details of the tracked mentions and distribution of frequent hashtags and bigrams in the data are available in \cite{chen2020election2020}.

For this analysis, we focus on the tweets that appeared between June 20, 2020 and September 6, 2020, in order to study the disinformation surfaced on Twitter in the months preceding the election. This subset of the data contains $242,087,331$ election-related tweets from $10,392,492$ unique users. We focus on engagements, therefore, it is useful to define four tweet types considered here: (i) Original tweets (accounts can create content and post on Twitter) (ii) reply tweets (iii) retweeted tweets, which reshare without comment (iv) quote tweets (embed a tweet i.e., reshare with comment). Engagements are defined here as one of the tweet types other than original tweets (and engagements can be with original tweets or other tweet types also).

\section{Disinformation Detection Methodology}
\label{csi}

\begin{table}[t]
    \centering
    \caption{Detection dataset statistics}
    \label{tab:detection_datastats}
    \resizebox{0.45\textwidth}{!}{
    \begin{tabular}{l|r}
        \toprule
         Statistic &  Count \\
         \midrule
         Tweets & 242,087,331
         \\ 
         Users accounts & 10,392,492 \\
         Unreliable/conspiracy URLs cascades & 3,162 \\
         Reliable URLs cascades & 4,320 \\
         Unlabeled cascades & 192,103 \\
          Avg. cascade size (\# engagements) & 57.11 \\
          Avg. cascade time (in hrs) & 80.42 \\
         Avg. time between engagements (hrs) & 5.93 \\
         \bottomrule
    \end{tabular}
    }
\end{table}

\begin{table*}[t]
    \centering
    \caption{Results on detection of unreliable/conspiracy cascades in the election dataset}
    \label{tab:detection_results}
    \renewcommand*{\arraystretch}{1} \resizebox{0.85\textwidth}{!}{
    \begin{tabular}{c|c|c|c|c|c|c}
    \toprule
    Method & AUC & AP & F1 & Prec & Rec & Macro-F1 \\
    \midrule
    SVM & 0.6236 $\pm$ 0.01 & 0.5025 $\pm$ 0.01 & 0.5710 $\pm$ 0.01 & 0.5594 $\pm$ 0.01 & 0.5835 $\pm$ 0.02 & 0.6226 $\pm$ 0.01  \\
    GRU & 0.5244 $\pm$ 0.05 & 0.4499 $\pm$ 0.04 & 0.4606 $\pm$ 0.11 & 0.4367 $\pm$ 0.05 & 0.5110 $\pm$ 0.17 & 0.5073 $\pm$ 0.05 \\
    CI & 0.6326 $\pm$ 0.02 & 0.5364 $\pm$ 0.01 & 0.5732 $\pm$ 0.02 & 0.5307 $\pm$ 0.02 & 0.6243 $\pm$ 0.03 & 0.6046 $\pm$ 0.02 \\
    CI-t & 0.6554 $\pm$ 0.02 & 0.5661 $\pm$ 0.01 & 0.5820 $\pm$ 0.03 & 0.5354 $\pm$ 0.01 & 0.6426 $\pm$ 0.08 & 0.6099 $\pm$ 0.01  \\
    \textbf{CSI} & 
    \textbf{0.8054 $\pm$ 0.02} &  \textbf{0.6826 $\pm$ 0.02} &  \textbf{0.7597 $\pm$ 0.02} &  \textbf{0.6611 $\pm$ 0.02} &  \textbf{0.8944 $\pm$ 0.04} &  \textbf{0.7608 $\pm$ 0.02} \\ 
    \bottomrule
    \end{tabular}
    }
\end{table*}

In this section, 
we present the methodology to separate factual from unreliable (or conspiratorial) claims in election-related tweets.  Conspiracies theories are attempts to explain events or situations, often postulated on most likely false and unverifiable theories, and may be politically motivated  \citep{ferrara2020characterizing,phadke2021makes}. 
To capture distortion aimed at manipulation of public opinion, we focus on any kind of unreliable (likely false or misleading) narratives, including conspiracies promoted prior to the election. The term disinformation is used here as an umbrella term to refer to distorted narratives appearing in the form of unreliable, including conspiratorial claims. In literature, the term disinformation separates itself from misinformation based on intent to manipulate vs. lack thereof \citep{sharma2019combating}. However, since we do not assess intent from tweets, we do not make a distinction between the two terms.
 
\subsection{Model Specifications}

\paragraph{Detection model.} We apply CSI, a supervised deep learning based model in \cite{ruchansky2017csi} for disinformation detection. The model captures three main signals or features: source, text, temporal, which are useful for disinformation detection \cite{ruchansky2017csi,sharma2019combating}. To classify a tweet, the model considers time-ordered sequence of engagements that the tweet receives, along with temporal information such as the frequency and time intervals between engagements. In addition, it learns the source characteristic, based on account behaviours i.e., which accounts co-engage with a given tweet. These input signals are useful for differentiating disinformation tweets. In general, engagements or social context are useful to learn predictive features, as they provide feedback from other accounts through engagement dynamics \cite{sharma2019combating}. The model needs to be provided training data containing labeled cascades (i.e., tweets with its engagements), We describe the process for labeling data and extracting  features (cascades) for training the detection model in the following paragraphs.

\paragraph{Reliable, unreliable and conspiracy URLs.} 
We collect lists of unreliable and conspiracy news sources from three fact-checking resources on low-credibility news sources: Media Bias/Fact\footnote{\href{https://mediabiasfactcheck.com/}{https://mediabiasfactcheck.com/}}, NewsGuard\footnote{\href{https://www.newsguardtech.com/covid-19-resources/}{https://www.newsguardtech.com/covid-19-resources/}}, and \citeauthor{zimdars2016false} (\citeyear{zimdars2016false}). 
NewsGuard maintains a repository of news publishing sources that have actively published false information about the recent COVID-19 pandemic. The listed sources from NewsGuard, accessed on September 22, 2020 are included, along with low and very low factual sources listed as questionable from Media Bias/Fact Check, and sources tagged with unreliable or related labels from Zimdar's list. We separately collect mainstream reliable news sources referencing Wikipedia\footnote{\href{https://en.wikipedia.org/wiki/Wikipedia:Reliable_sources/Perennial_sources}{\url{https://en.wikipedia.org/wiki/Wikipedia:Reliable_sources/Perennial_sources}}
}. In total, we obtained 124 mainstream reliable and 1380 unreliable (or conspiracy) news sources. Original tweets sharing URLs published from these news sources are thereby labeled as reliable and unreliable/conspiracy respectively. In addition, we also label tweets that are retweets, replies, or quotes sharing such URLs, if the parent of the tweet is not in the collected tweets (1\% Twitter sample).



\paragraph{Information cascades and cascade statistics.} To train the detection model, we first extract \emph{cascades} of engagements from the dataset. Information cascades represent the diffusion or spread
of information on the network. Each cascade corresponds to a time-ordered sequence of engagements (retweets, quotes and replies), that originate at a source tweet and spread through chains of retweets, quotes and replies. Formally, a cascade can be denoted as $C_j = [(u_1, tw_1, t_1), (u_2, tw_2, t_2), \cdots (u_n,tw_n, t_n)]$, where the tuple corresponds to an original tweet, or an engagement with the original tweet or its predecessor engagements, and represents the user (u), tweet (tw), and temporal (t) dimensions of when the user posted the tweet or engagement. The cascade is labeled by the URLs linked in its first tweet as described earlier, if the tweet shares a URL from one of the considered factual or unreliable news sources, it is labeled accordingly, otherwise left unlabeled.

Since the dataset contains over 10M accounts and 242M tweets, for computational efficiency, we subsample accounts and cascades, whilst ensuring minimal information loss, such that at least 75\%($\sim$180M) of the tweets are accounted for in the data. We find that this can be achieved by subsampling cascades with a minimum number of engagements, and accounts with highest active and passive engagements with other accounts in the dataset\footnote{Active refers to tweets (original, retweet, reply, quote) from the account, whereas passive means the account has been involved in another account's tweets (retweeted, replied, quoted, or mentioned)}. For accounting for $>$75\% of the tweets, we find cascade size of 5, and 7471 highest engaging accounts was sufficient. This results in $\sim$200K cascades (192K unlabeled, and rest labeled with 3120 unreliable and 4320 reliable cascades). The full cascades comprise engagements from accounts outside of the subsampled accounts, and span the large fraction of dataset tweets. Full cascades are used in the analysis of identified disinformation cascades, but for training and inference with CSI, the subsampled accounts in these cascades are utilized. 
Table~\ref{tab:detection_datastats} provides statistics of data and training cascades. 

\subsection{Model Evaluation}

Table~\ref{tab:detection_results} reports 5-fold cross validation results obtained using CSI on the labeled cascades. The metrics reported are the ROC AUC average precision AP score. We also reported the F1, Precision, Recall and macro-F1 at the detection threshold selected from the ROC curve, that achieves maximum geometric mean of validation set sensitivity and specificity. 

For comparison, we evaluate it against several baselines: \textit{(i)} SVM- RBF on text features of the cascade source tweet, extracted with doc2vec. \textit{(ii)} GRU \citep{ma2016detecting} which utilizes a recurrent neural network (RNN) to classify the cascade based on text features of the time-ordered sequence of engagements in the cascade. \textit{(iii)} CI, which can be considered as a variant of the CSI model, utilizes only tweet text and account metadata features of the engagements as input to a RNN for classification.
\textit{(iv)} CI-t, another variant of the CSI model, utilizes tweet features of engagements, and in addition temporal information of time intervals between engagements as input to a RNN. (v) CSI model, which includes all three features (tweet features of engagements, temporal information, and account behaviours). 

The CSI model has AUC 0.8 and high recall and is therefore used for inference on unlabeled cascades. The ensemble of CSI models trained over five folds are used for inference. We take the top 80th and bottom 80\% percentile, of cascades \emph{predicted with the largest margins} above and below the detection threshold, resulting in 72,228 unreliable and 81,453 reliable classified cascades respectively. The engagements or tweets forming the reliable cascades constitute about two third of the total tweets in all cascades. 
\update{For human
validation, we take a random sample of 50 cascades from ones labeled by the model (Note: the list of validated tweets is provided in supplementary materials). In 2/50 (4\%) the model label differed from the label assigned by inspecting the tweet and its account for suspensions. The model errors included a tweet critical of the postal system and though non conspiratorial, could have been mistaken for claims related to mail-in voter fraud by the model, and another which strongly supported criticism of China by president Trump, although not from a suspended account and unlikely part of QAnon/conspiracy groups, was predicted unreliable. To reiterate, these validated cascades were uniformly sampled from cascades that were assigned labels using CSI, considering predictions with largest margins from the detection threshold, as discussed earlier using the ensemble of 5-fold classifiers, and the expected lower error rate, confirmed by validation suggests it can be reliably utilized for further analysis.}




\section{Methodology for Inferring Political Leaning}

In this section, we describe the methodology we use to infer political leaning of accounts on the network. The inferred political leanings are used later in the analysis and characterization of the disinformation landscape. Similar to prior work \citep{ferrara2020characterizing}, we use the list of 29 prominent news outlets classified as left, lean left, center, lean right, right as per ratings provided by allsides.com.\footnote{\href{https://www.allsides.com/media-bias/media-bias-ratings}{\url{https://www.allsides.com/media-bias}}} We consider left and lean left classified outlets as the left-leaning, and right and lean right as right-leaning.

\subsection{Model specifications}
For accounts that tweeted or retweeted URLs published from these news outlets, we characterize their political leaning by measuring the average bias of the media outlets they endorsed. This gives us a set of labeled accounts with known political leaning labels based on the media URLs. To infer labels of other accounts, we can propagate labels from this known set (also called the seed set i.e., accounts with known political leaning labels based on the media URLs) to other accounts based on interactions between accounts. 

We utilize the retweet network to infer political leaning of other accounts starting from the seed set accounts. Retweeting is a form of endorsement, different from other tweet types, and accounts which retweet each other tend to share the same political biases  \citep{badawy2019characterizing}. To that end, we use \emph{Louvain} method \citep{blondel2008fast} to identify communities in the retweet graph, where edge weights represent the number of retweets between accounts. The method optimizes modularity which provides a measure of edge density within communities compared to across communities. We assign political leaning to each identified community, using the average political leaning of media URLs endorsed by accounts in the seed set that belong to the community. The seed set accounts with high entropy in distribution of left-leaning and right-leaning URLs (close to uniform distribution with margin of 0.2), and ones that shared less than 10 URLs from the media outlets are filtered out. 

\subsection{Inference of political leanings} Using media outlets, we obtain a seed set of 114K accounts from 10.4M accounts in the dataset. To limit the size of the retweet network, we consider the top active 1.2M accounts that appeared (in original tweet, retweet, quote, or reply tweet) at least 20 times in the collected dataset. 

Using the seed set and retweet graph of 1.2M accounts, gives a resulting inferred network with large left-leaning communities of 540,719 and 68,197 accounts and two smaller left-leaning ones, and large right-leaning communities of 480,982 and 10,723 accounts, and multiple smaller ones. Of the 1.2M accounts, we were able to infer the political leaning of 92\% of the accounts. The rest of the accounts remain undetermined due to high entropy in left and right-leaning URLs shared, or with communities that had fewer than two seed accounts. We thereby identified 610,430 left-leaning and 500,804 right-leaning in the 1.2M accounts. 

\update{\textbf{Verification}. We measure the accuracy of inferred leanings based on three types of evaluations (i) Media URL labels i.e., based on the averaged political leaning of left/right leaning media outlets endorsed in tweets from the account (However since media URL labels are also used as seed set labels during inference, therefore for evaluation we report averaged 5-fold results wherein 20\% of the seed labels are held-out and kept unseen during inference). (ii) Profile descriptions, i.e., based on whether the account profile description dominantly included left or right leaning hashtags (the hashtags were classified as left/right through human validation of most frequently used 3,000 hashtags. The list is provided in supplementary material). (iii) Manual verification based on inspection of tweets of randomly sampled subset of accounts (i.e., based on explicitly stated party affiliation in tweet or account profile, or expressed support of left/right presidential candidate/party, and assigned `center' instead of left/right if non-partisan). We sampled 100 inferred accounts uniformly at random, and 124 inferred accounts by stratified sampling based on degree distribution of accounts in the retweet graph, to ensure coverage of dense and sparsely connected accounts in the retweet graph, since the retweet graph was used infer the leanings.}

\update{In Table~\ref{tab:political_inference_validation}, we report the error rate on each evaluation measure separately for the left-leaning and right-leaning accounts. Here, we added an alternative baseline based on Label propagation \cite{badawy2019characterizing} for comparison. The total error rate was 4.46\% on manually verified labels for both methods (label propagation (LP) and Louvain (Lo) used here, on RT graph from same media URL labeled seed set) and results were robust on both. Error analysis on the manually labeled set, suggests that errors included few accounts that are actually neither left/right leaning (but center), e.g., U.S. Department of State, reporter accounts, or unrelated or disinterested in U.S. politics, and were erroneously classified as left/right leaning. Other mistakes included anti-Trump conservatives that were inferred as left-leaning, or with high entropy in left/right leaning views in their tweets.}

\begin{table}[t]
    \centering
    \caption{\update{Number of accounts labeled as left or right-leaning  (by media URLs, account profile description, and human verification) for validation, with error rate (\%) in each type based on the inferred political leaning of those accounts.}} 
    \label{tab:political_inference_validation}
    \begin{tabular}{l|c|c|c}
    & MediaURLs & Profile Desc. & Human Verif. \\
    \toprule
    LP-Left & 68k (0.71) & 29.5k (0.32) & 116 (1.72) \\
    LP-Right & 46k (0.27) & 14.0k (0.34) & 103 (2.91) \\
    \midrule
    Lo-Left & 68k (0.67) & 29.5k (0.25) & 116 (1.72) \\
    Lo-Right & 46k (0.37) & 14.0k (0.58) & 103 (2.91) \\
    \bottomrule
    \end{tabular}
\end{table}


\section{Results and Analysis}

\begin{figure*}[t]
    \centering
    \includegraphics[width=17cm,height=8.5cm]{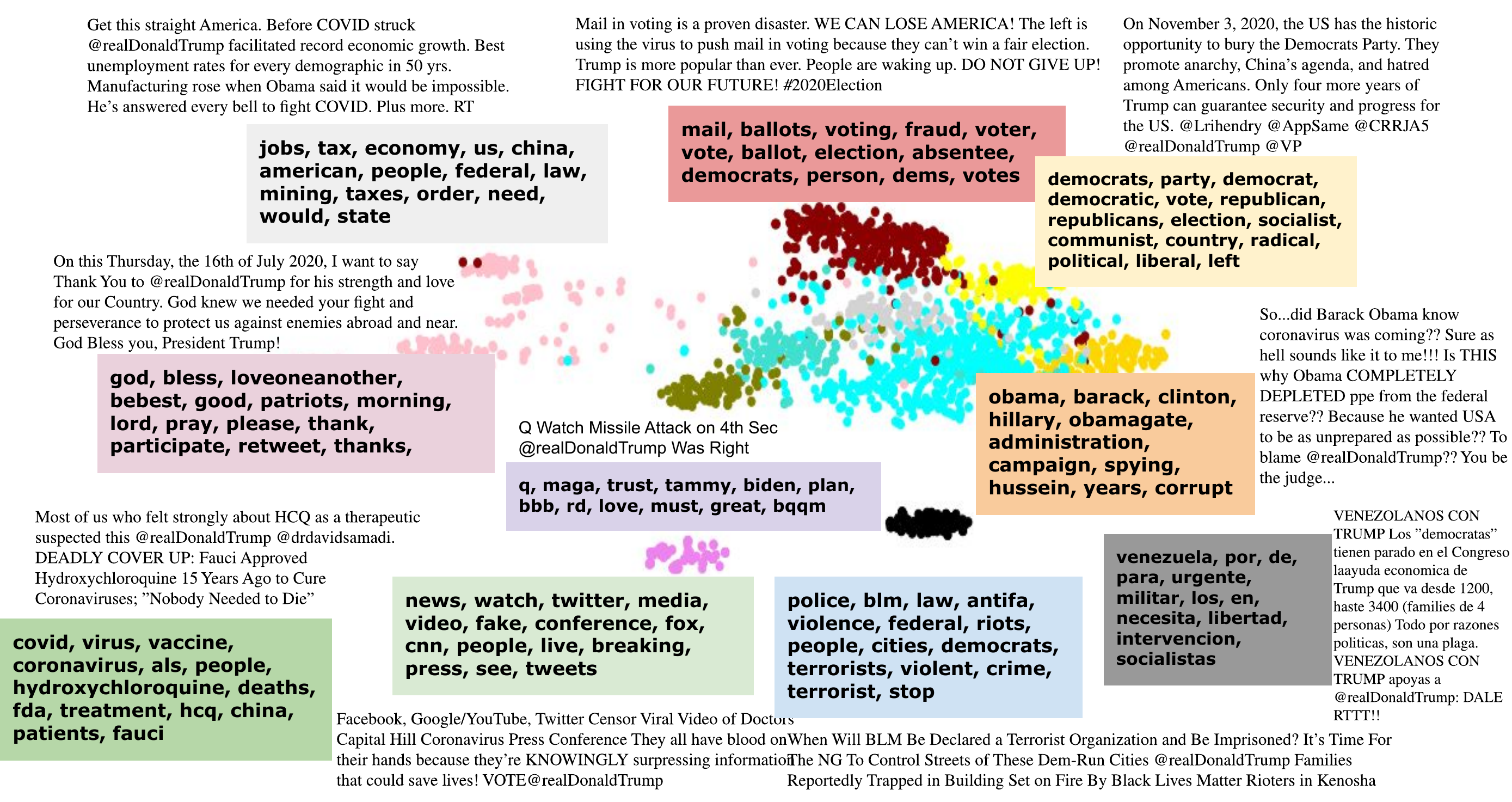}
    \caption{Topic clusters for identified unreliable/ conspiracy tweets with example top representative tweet of cluster}
    \label{fig:topics}
\end{figure*}

In  the  following subsections,  we analyze identified  unreliable/conspiracy  cascades,  and characterize account engagements. We characterize disinformation topics, QAnon conspiracy  group  and  its  interactions  with left  and  right demographics, their activities, effect of Twitter’s restrictions, and propagation dynamics of engagements with disinformation.

\subsection{Disinformation topic modeling} 
\paragraph{Topic modeling} We use topic modeling to identify prominent topics that were targets of disinformation prior to the election. With identified unreliable/conspiracy cascades, we can model the topics in tweet text associated with the source (first) tweet in the cascade. The text is pre-processed by tokenization, punctuation removal, stop-word removal, and removal of URLs, hashtags, mentions, and special characters, and represented using pre-trained fastText word embeddings \citep{bojanowski-etal-2017-enriching}\footnote{Pre-trained: \href{https://fasttext.cc/docs/en/english-vectors.html}{\url{https://fasttext.cc/docs/en/english-vectors.html}}}. We take the average of word embeddings in the tweet text to represent each tweet. Using pre-trained embeddings trained on large English corpora, we can potentially encode more semantic information and it is useful for short texts where word co-occurrence statistics are limited for utilizing traditional probabilistic topic models \citep{li2016topic}. The tweet text representations are clustered using k-means to identify topics clusters. 

We select number of clusters (K) using silhouette and Davies-Bouldin measures of cluster separability. K that jointly is in the best scores of both is selected between 3-35. This gave us K=30, and inspecting word distribution and representative tweets (closest to cluster center),
we discard two clusters unrelated to US politics (ALS treatment and Nigeria violence), and eight small or less distinguished clusters, and merge over-partitioned clusters each related to the Black Lives protests, and to mail-in voter fraud.

\paragraph{Topic clusters} The resultant clusters are in Fig~\ref{fig:topics}. For each cluster, top words ordered by highest tf-idf scores, along with an example tweet from 100 most representative tweets of the cluster are shown. The major themes relate to false claims about mail-in voter fraud, COVID-19 and pushing hydroxychloroquine as a cure, and protests concerning law enforcement and Black Lives Matter. Other topics target specific candidates and entities, such as social media platforms for censorship of unverified and conspiratorial content, conspiracies and allegations against former president Obama, or targeting the democratic party as a whole on different social issues, and misleading claims about jobs and economy. The remaining clusters include the QAnon conspiracies, a far-right conspiracy group now banned by several platforms \citep{de2020tracing}. Another cluster related to Venezuela appears in support of right-leaning theories potentially about voter fraud, and anti-democratic posts, however, our modeling is limited to English, the most prominent language in the tweets ($\sim$94\% disinformation tweets were in English, followed by $\sim$3\% in Spanish, remaining languages less than 0.1\% each). 
In Table~\ref{tab:most_engaged_disinformation}, we list examples of identified unreliable/conspiracy tweets within tweets with the most engagements in the collected dataset, discarding false positives. Some of these have been debunked by fact-checking sites as false, misleading or lacking evidence\footnote{\href{https://www.politifact.com/factchecks/2020/jul/28/viral-image/opening-case-file-does-not-mean-joe-biden-criminal/}{\url{https://www.politifact.com/factchecks/2020/jul/28/viral-image/opening-case-file-does-not-mean-joe-biden-criminal/}}}\footnote{\href{https://apnews.com/article/shootings-wisconsin-race-and-ethnicity-politics-lifestyle-23668668b3b59fa609a18d023c0bb485}{\url{https://apnews.com/article/shootings-wisconsin-race-and-ethnicity-politics-lifestyle-23668668b3b59fa609a18d023c0bb485}}}. 


\begin{table}[t]
    \centering
    \caption{Examples of unreliable/conspiracy tweets within tweets with most engagements in the data}
    \label{tab:most_engaged_disinformation}
    \resizebox{.5\textwidth}{!}{
    \begin{tabular}{p{10cm}|p{1cm}}
    \toprule
 Unreliable/conspiracy tweet & \# Eng. \\
 \midrule
 BREAKING: Democratic Presidential nominee @JoeBiden is formally being listed as a criminal suspect by high level Ukraine government officials, in a major case involving his son - Hunter. https://t.co/Xe2bSLEAh8 & 56K \\
 \midrule
 We are being censored! @realDonaldTrump @Facebook is about to unpublished our FB page with 6 million followers. The NY Times recent article claiming we are right wing Provacateurs They are interfering with this election! Conservatives are being censored on FB. PLEASE RETWEET!! https://t.co/xVy8xZ7kyC & 48.6K \\
 \midrule
 I am tired of the censorship! Anderson Cooper called me a snake oil salesman because I’m trying to get the FDA to test a supplement that I've seen work! And Twitter keeps taking my followers! Please RT and follow me as I support President @realDonaldTrump! & 48K \\
 \midrule
 Can you believe what’s happening!? They give Joe Hiden’ the questions, and he reads them an answer! https://t.co/ivMw6uQ2gp & 47K \\
 \midrule
 HEARTBREAKING A 60 year old Black Trump Supporter was murdered in cold blood all because he support President @realDonaldTrump This is a Hate Crime He deserves Justice Let’s make his name trend Use \#JusticeForBernellTrammell https://t.co/XZbdOiHgRR & 46.9K \\
 \midrule 
 NATURE ARTICLE HOAX BUSTED!! Proof that chloroquine let’s covid attack cancer cells but not normal cells. PLEASE RETWEET. @realDonaldTrump @IngrahamAngle @SteveFDA @drsimonegold @jennybethm https://t.co/XN0YC1liSQ & 40.6K \\
 \midrule
 If we can stand in line at a grocery store or hardware store, we can stand in line at the polls to vote. President @realDonaldTrump is RIGHT that universal, unmonitored vote-by-mail would be a DISASTER, and we’re already seeing evidence of that across the country. @TeamTrump https://t.co/ai1uNjQi7k & 32.5K \\
\bottomrule
    \end{tabular}
    }
\end{table}

\subsection{Quantifying interactions with QAnon group}

\begin{table}[t]
    \centering
    \caption{QAnon conspiracy keywords along with their occurrence frequency in tweets (original, reply or quoted tweets i.e., excluding retweets) containing the keywords} 
    \label{tab:Qanon_keywords}
    \resizebox{0.4\textwidth}{!}{
    \begin{tabular}{l|l|l|l}
    \toprule
    Keyword & Freq. & 
    Keyword & Freq. \\
    \midrule
        wwg1wga & 159436 &
        wgaworldwide & 18231 \\
        \#qanon & 68039 &
        \#qarmy & 13577  \\ 
        \#obamagate & 78574 &
        \#pizzagate & 13053 \\
       \#savethechildren & 33221 & 
       \#taketheoath & 10994 \\
         thegreatawakening & 23305 &
        greatawakening & 31615 \\
        deepstate & 25268 &
        deepstatecoup & 995 \\ 
        deepstatecabal & 1669 &
      deepstateexposed & 2188 \\
      \#pedogate & 5211 &
        pedowood & 4454 \\
      \#plandemic & 8456 &
      \#scamdemic & 4674 \\
      \#sheepnomore & 1492 &
      adrenochrome & 6397 \\
      thestorm & 3989 &
      followthewhiterabbit & 95 \\
      thesepeoplearesick & 2843 &
      wearethenewsnow & 5540 \\
      trusttheplan & 2579 &
      pizzagateisreal & 698 \\
      thestormisuponus & 887 &
      newworldorder & 901 \\
      darktolight & 6825 &
      clintonbodycount & 1898\\
    \bottomrule
    \end{tabular}
    }
\end{table}

\begin{table}[t]
    \centering
    \caption{\update{Verification of 100 accounts sampled from inferred right/left-leaning accounts posting QAnon associated keywords. Verified as: (Q) is QAnon conspirator; else not (N)}}
\label{tab:qanon_validation}
    \resizebox{0.45\textwidth}{!}{
    \begin{tabular}{l|l|c}
    \toprule
Inferred Leaning & Human verification & \# Accounts \\
\midrule
Right-leaning (74k) & Q & 100 \\
Left-leaning (7.6k) & N (Reference QAnon) & 79 \\
 & N (Re-purposed hashtag) &  2 \\
& Q (Incorrect leaning) & 19  \\
Undetermined (10k) & Q & 89 \\
& N & 11 \\
\bottomrule
\end{tabular}
}
\end{table}

\begin{figure}[t]
    \centering
    \includegraphics[scale=0.6]{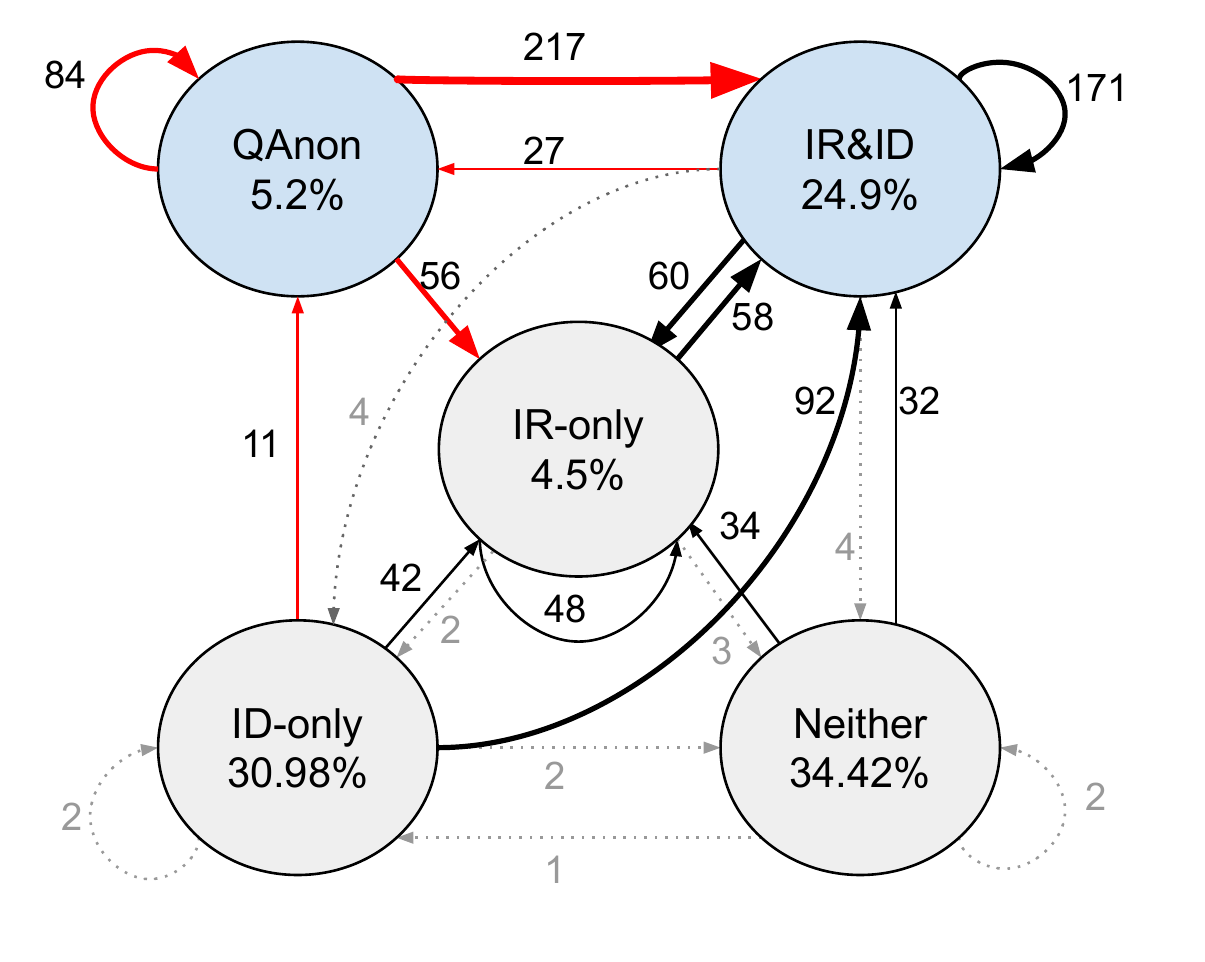}
    \caption{QAnon accounts interaction graph in active 1.2M accounts. Edges are retweets/quotes/replies from source to destination node, normalized by \# accounts in source.}
    \label{fig:Qanon_interaction_graph}
\end{figure}

\begin{table}[t]
    \centering
    \caption{QAnon interactions quantified over all accounts in the dataset. Influenced (ID) are accounts that replied, retweeted or quoted tweets from QAnon accounts. Influencer (IR) are accounts that were replied, retweeted or quoted by QAnon accounts. Table contains: \# Accounts (\%).}
    \label{tab:QAnon_interaction_stats}
    \resizebox{0.5\textwidth}{!}{
    \begin{tabular}{l|r|r|r}
    \toprule
     & 10.3 M (All users) & \multicolumn{2}{c}{1.2M (Activity $>$ 20)} \\
    \midrule
    Group & QAnon 74.3k (0.72) & QAnon 62.8k (5.2) & \update{Sample} \\
    \midrule
    IR\&ID & 376k (3.62) &  300.7k (24.9) & \update{(22.60)} \\
    IR-only & 150.6k (1.45) & 54.3k (4.5) & \update{(3.00)} \\
    ID-only & 912.5k (8.78) & 374k (30.98) & \update{(47.01)} \\
    Neither & 8.9M (85.43)	& 415.7k (34.42) & \update{(22.19)} \\
    \bottomrule
    \end{tabular}
    }
\end{table}

\begin{figure}[t]
    \centering
    \caption{Distribution of account interactions with QAnon accounts by tweet type and inferred political leaning.}
    \label{QAnon_tt_political_leaning}
    \begin{subfigure}{0.5\textwidth}
    \centering
     \caption{\% of accounts that interacted with QAnon accounts with mentioned tweet types inferred as left or right leaning with active 1.2M accounts.}
     \label{QAnon_tt_political_leaning:tweet_type}
     \resizebox{0.85\textwidth}{!}{
        \begin{tabular}{l|p{1cm}|p{1cm}|p{2cm}}
        \toprule
           Tweet Type & Left-Leaning &  Right-Leaning & Undetermined \\
        \midrule
            Reply To (RP-TO) & 42.54 & 50.16 & 7.30 \\
            Replied By (RP-BY) & 49.97 & 45.58 & 4.45 \\
            Retweet (RT) & 8.95 & 90.06 & 0.99 \\
            Quoted (QTD) & 25.03 & 73.44 & 1.53 \\
        \bottomrule
        \end{tabular}
        }
    \end{subfigure}
    ~
    \begin{subfigure}{.5\textwidth}
    \vspace{2em}
    \centering
    \caption{Fraction of inferred left/right-leaning accounts in active 1.2M that interacted with QAnon accounts with mentioned tweet types.}
    \label{QAnon_tt_political_leaning:political_leaning}
        \includegraphics[width=8cm,height=3.5cm]{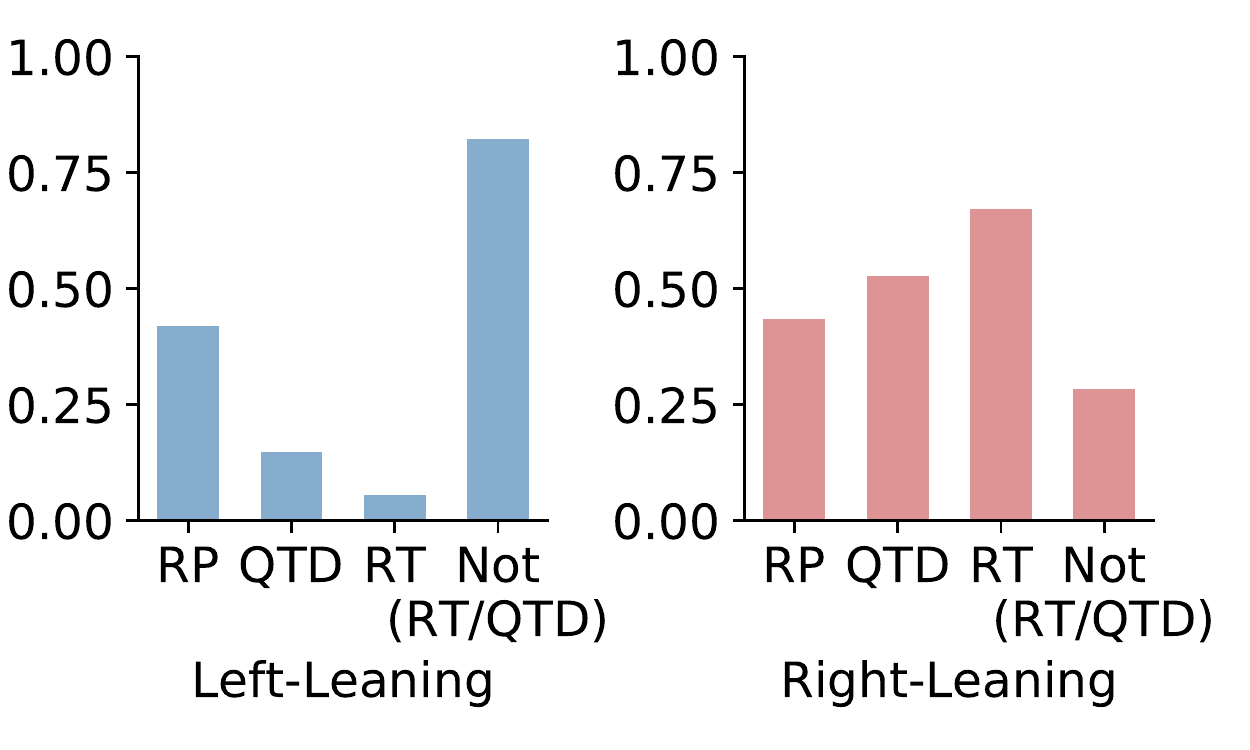}
    \end{subfigure}
\end{figure}

\begin{figure*}[t]
    \centering
    \begin{subfigure}{0.50\textwidth}
    \centering
    \includegraphics[width=\textwidth,height=4cm]{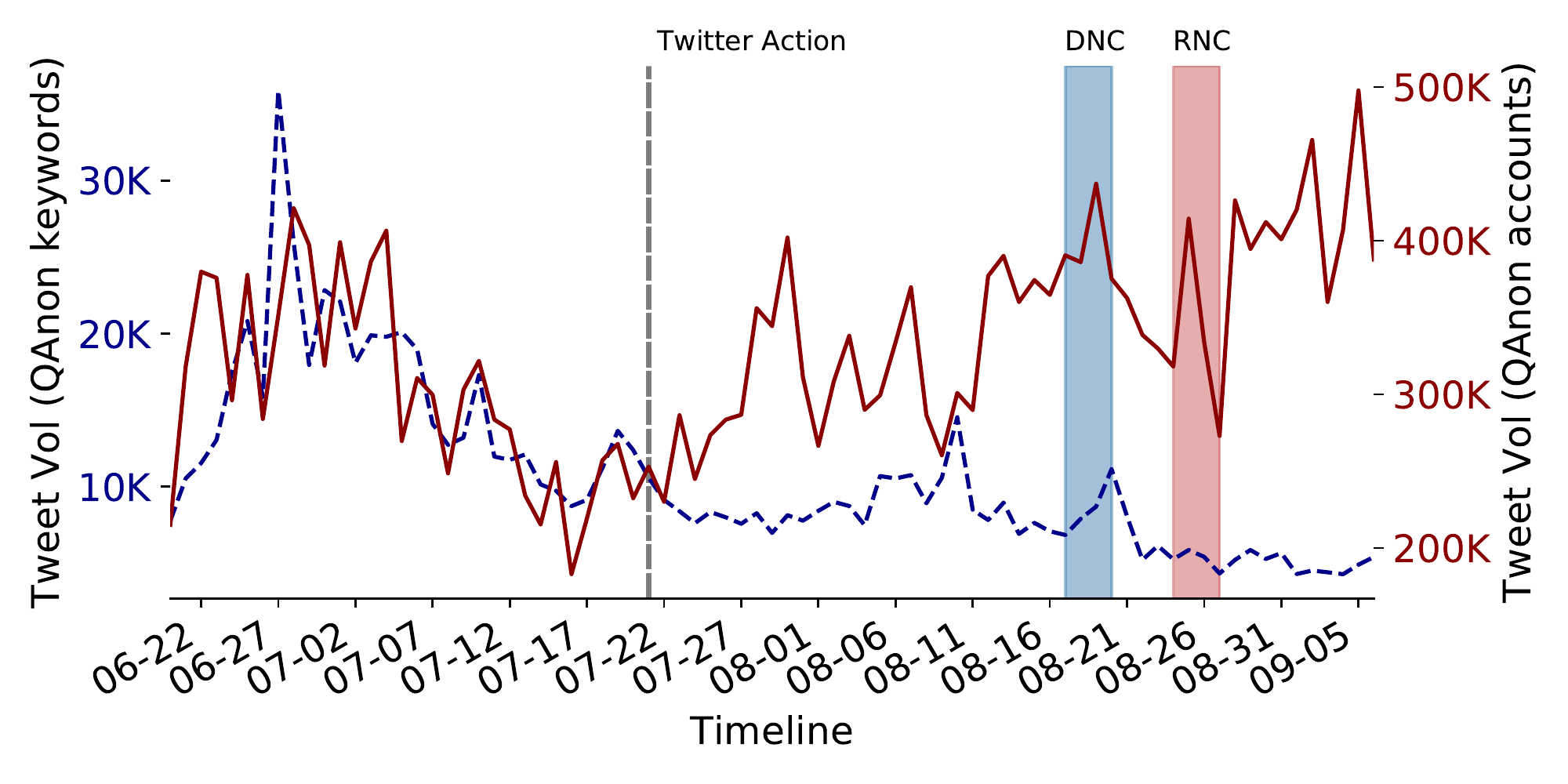}
    \caption{QAnon tweets activity timeline. (Red) Total daily volume of tweets from the QAnon accounts in the data collection period. (Blue) Daily volume of QAnon account tweets containing the listed QAnon keywords.}
    \label{fig:Qanon_activity_timeline}
    \end{subfigure}
    ~
    \begin{subfigure}{0.48\textwidth}
    \centering
    \includegraphics[width=\textwidth,height=3.8cm]{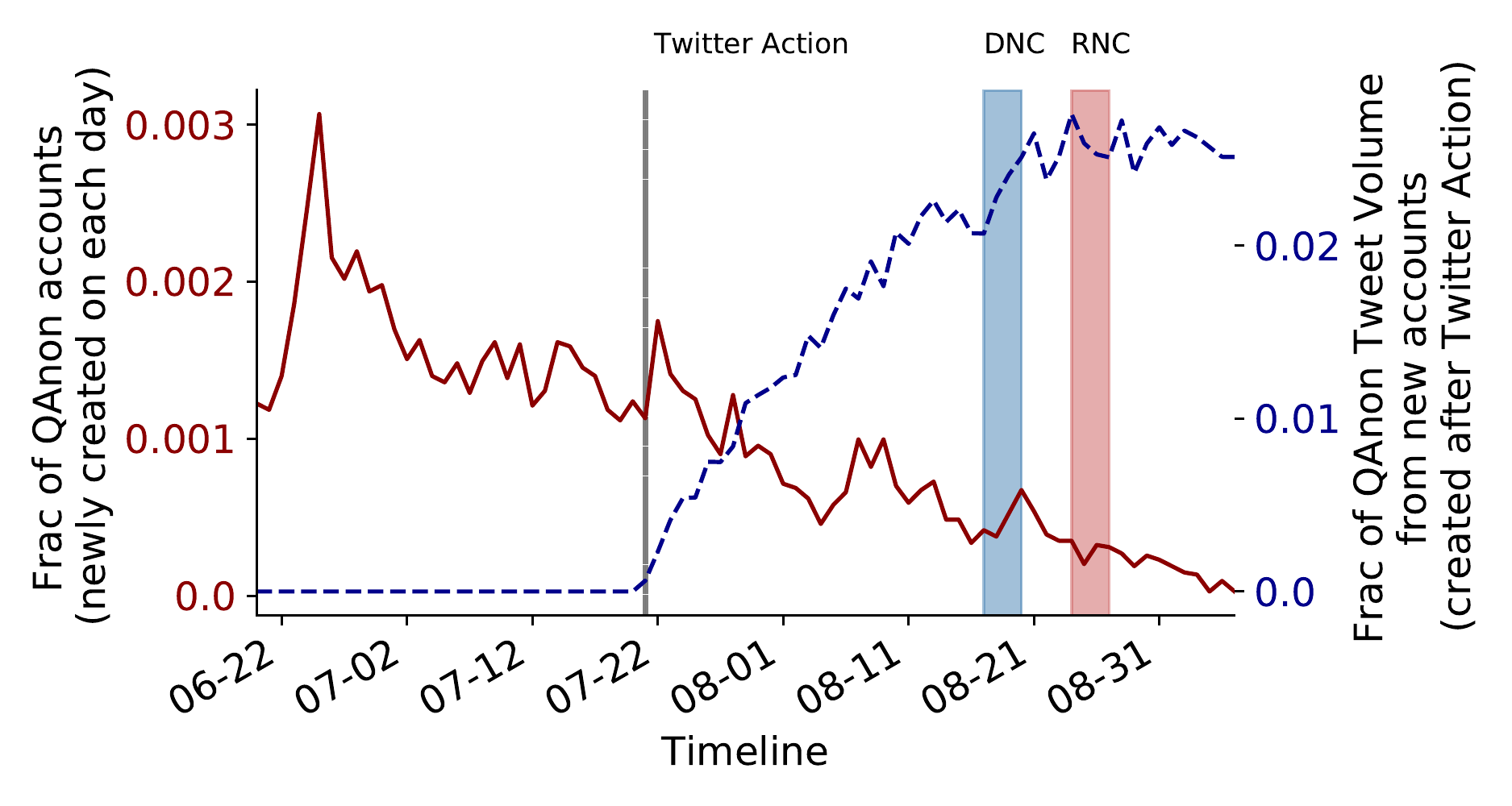}
    \caption{QAnon accounts creation timeline. (Red) Fraction of newly created QAnon accounts per day in the data collection period. (Bottom) Fraction of total tweet volume from QAnon accounts that is  attributed to new accounts created after Twitter Action.}
    \label{fig:Qanon_account_creation_timeline}
    \end{subfigure}
    \caption{QAnon activity and account creation timeline with respect to Twitter Action banning QAnon, DNC and RNC.}
    \label{fig:Qanon_timeline}
\end{figure*}

QAnon, a far-right conspiracy group emerged in 2017 on 4chan, 
and has risen to prominence for its baseless conspiracies that have received significant following and attention \citep{de2020tracing}. The group advances the conspiracy theory that president Trump is battling a satanic child sex-trafficking ring, and an anonymous `Q' claims to be a US government official with top-clearance, providing insider information about deep state operations \citep{ferrara2020characterizing}. In this subsection, we analyze activities of QAnon accounts and characterize its interactions with other accounts.

\textbf{ Identification and verification.} We identify accounts that actively form part of the conspiracy group by posting original content related to QAnon conspiracies, referred to as QAnon accounts thereafter. We extract tweets (original, quote, or reply tweets) excluding retweets containing any keywords or hashtags frequently associated with the QAnon group. Table~\ref{tab:Qanon_keywords} lists the QAnon associated keywords with their frequencies. This gives 92,065 accounts
with posts containing QAnon associated keywords. 7,661 of these were inferred left-leaning accounts, for 10,085 the political leaning was undetermined (not inferred), and the rest were inferred as right-leaning. 
\update{For accounts posting QAnon keywords, grouped by the inferred political leaning, we sampled 100 accounts uniformly at random from each group (left, right, undetermined), and inspected their tweets to identify whether the accounts are promoters of QAnon conspiracies. The validation and error analysis of the sampled accounts is provided in Table~\ref{tab:qanon_validation}. Among the 100 inferred left accounts, 19\% were mistakenly inferred as left-leaning (and were promoters of the conspiracy), 79\% were not part of the conspiracy group and only referencing the QAnon movement (oppose, ridicule, discuss or tag in their tweets containing QAnon keywords), and 2\% were using the ``pedogate" hashtag in a different context as compared to how the hashtag was used by QAnon conspirators. Among the inferred right-leaning accounts, all were verified to be QAnon accounts.}

\update{Based on this observation, we retained the inferred right-leaning accounts posting QAnon associated keywords as the identified QAnon accounts (since they indicate  with high precision support of the far-right QAnon conspiracy theories). This two-stage filtering for QAnon accounts identification, results in 100\% \emph{precision} and 88\% \emph{recall} based on the verified examples weighted by actual number of accounts in each type (words not included as QAnon associated keywords are not considered in the recall measure).} 

\paragraph{QAnon interaction graph.} To quantify interactions between accounts, we consider that two accounts have interacted if there exists a retweet (i.e., retweet without comment), quoted tweet (i.e., retweet with comment), or reply tweet edge between the accounts, in the collected dataset. Further, to account for directionality of the interactions, we partition the accounts that have interacted with QAnon accounts into ``ID" (influenced) and ``IR" (influencer). Influenced accounts are those that have been influenced by (retweeted, quoted, or replied to) QAnon accounts tweets, whereas influencer are accounts that QAnon accounts have retweeted, quoted or replied to. Table~\ref{tab:QAnon_interaction_stats} reports what percentage of all accounts were influenced but did not influence QAnon (ID-only), influencer but not influenced (IR-only), and both influencer and influenced in the data (IR\&ID).

We find that 85.43\% of 10.4M accounts had no observed interaction with QAnon accounts in the collected tweets. However, considering accounts that appeared in at least 20 tweets (original, reply, retweet, quote) in the dataset, we find only 34.42\% of the active 1.2M accounts had no interaction with QAnon accounts in the collected tweets. \update{Also, we consider a random sample of accounts from the active 1.2M accounts of the same sample size as \% of QAnon accounts in that set (drawn five times, and statistics averaged) for comparison.  IR\&ID  and  IR-only are similar. The main difference is in the ID-only and Neither cases, wherein \% of influenced only accounts by QAnon was less than the random samples (30.98 vs.  47.01), \% of accounts not interacting with QAnon is higher than with random samples (34.42 vs.  22.19).} 

The interaction graph of the 1.2M active accounts (activity $>$ 20 in the dataset) is visualized in Fig.~\ref{fig:Qanon_interaction_graph}. Thus, 5.2\% of these accounts are QAnon accounts, as described earlier. And 24.9\% (IR\&ID) are closely coupled accounts that both influenced QAnon content and were influenced by it (retweeted, quoted, replied). A larger 30.98\% were only influenced (ID-only), while a small fraction of 4.5\% were retweeted, quoted, or replied by QAnon, but did not engage back (IR-only). The edge weights in the interaction graph correspond to the volume of retweets, replies, or quoted tweets of the destination node by the source node, per account in the source node. 

\emph{QAnon engagement characteristics.} Next, we characterize interactions with QAnon accounts by inferred political leaning and tweet types, in the active 1.2M accounts in Fig.~\ref{QAnon_tt_political_leaning:tweet_type}. As one would expect, the retweeted and quoted tweet interactions are mostly between right-leaning and QAnon accounts (90.06\% and 73.44\%), as these are more close to endorsements. Replies on the other hand are roughly equally distributed amongst left and right-leaning accounts (49.97\% of replies by QAnon are towards left-leaning and 42.54\% of the replies to QAnon tweets are from left-leaning accounts).

This suggests that QAnon get engaged in active discussions with left-leaning accounts. This might also be indicative of engagement strategies by QAnon to influence liberals by ``red pilling", i.e., have their perspective transformed, illustrated in a QAnon account tweet which states:

\begin{quote}
``We interact with liberals the most under \@realDonaldTrump's most recent posts. Q gave us a challenge so I have been sharing the truths about COVID \& nursing home deaths in hopes of redpilling anyone who sees. Next Trump tweet wanna help out? https://breitbart.com/politics/2020/06/22/exclusive-seema-verma-cuomo-other-democrat-governors-coronavirus-nursing-home-policies-contradicted-federal-guidance/."

\end{quote}

Fig.~\ref{QAnon_tt_political_leaning:political_leaning} shows distribution of inferred left/right-leaning accounts that interacted with QAnon, by tweet types. 82.17\% of left-leaning in 1.2M have not retweeted/quoted QAnon accounts, but only 28.35\% of right-leaning have not endorsed content from QAnon accounts. 

\subsection{Estimating change in QAnon activities after Twitter Action}

Twitter announced sweeping actions against the QAnon conspiracy group on July 21, 2020, designating them as harmful coordinated activity. Twitter took down more than 7,000 QAnon accounts violating platform manipulation policies with spam, ban evasion, and operation of multiple accounts. Twitter also banned QAnon-related terms and urls from appearing in trending topics and search, limiting 150,000.\footnote{\href{https://www.nbcnews.com/tech/tech-news/twitter-bans-7-000-qanon-accounts-limits-150-000-others-n1234541}{\url{https://www.nbcnews.com/tech/tech-news/twitter-bans-7-000-qanon-accounts-limits-150-000-others-n1234541}}}

\paragraph{QAnon activities timeline.} Timeline of tweet volume (original, retweet, quote, reply) from QAnon accounts is visualized in Fig.~\ref{fig:Qanon_activity_timeline}. We separately plot the total daily volume of tweets and the daily volume of tweets containing the QAnon keywords (frequently associated with QAnon listed earlier in Table~\ref{tab:Qanon_keywords}). As observed, although the volume of tweets with QAnon associated keywords sharply declines after the Twitter Action, but the overall volume of activities from these accounts is sustained even after imposed bans and restrictions. This clearly suggests a gap in enforcement actions and desired outcomes. So we investigate evasion strategies that rendered Twitter Action ineffective.

In Fig.~\ref{fig:Qanon_account_creation_timeline} we inspect whether new QAnon accounts were injected to continue activities after the Twitter Action. Whilst a declining and small fraction of new accounts were introduced, the fraction of the total volume of QAnon account tweets  attributed to new accounts created after Twitter Action is less than 3\%. Clearly, much of the volume is sustained by earlier accounts even after the ban.

\paragraph{Evasion strategies with emerging and declining hashtags.}

\begin{table}[t]
    \centering
    \caption{Top 10 hashtags that \update{declined ($\beta < 0$)} in usage by QAnon post Twitter Action, estimated by regression discontinuity design in top 10K hashtags used by QAnon accounts}
    \label{tab:qanon_rdd_declining}
    \resizebox{0.45\textwidth}{!}{
    \begin{tabular}{l|c|c|c|c}
    \toprule
    \textbf{Hashtag} & $\bm{|\beta|}$ & slope & intercept & p-value \\
    \midrule 
wwg1wga & 1.512 & -0.047 & 4.951 & \textbf{0.002} \\
qanon & 0.901 & -0.03 & 2.974 & \textbf{0.007} \\ 
kag & 0.411 & -0.002 & 1.419 & \textbf{0.000} \\ 
q & 0.226 & -0.004 & 0.584 & \textbf{0.000} \\ 
qarmy & 0.207 & -0.008 & 0.7 & \textbf{0.048} \\ 
wwg1gwa & $0.139$ & -0.005 & 0.47 & 0.071 \\ 
qanons & $0.108$ & -0.004 & 0.42 & \textbf{0.031} \\ 
patriotstriketeam & $0.087$ & -0.001 & 0.164 & \textbf{0.010} \\ 
deepstate & 0.082 & -0.001 & 0.233 & \textbf{0.000}  \\ 
inittogether & 0.067 & -0.002 & 0.19 & 0.083 \\ 

    \bottomrule
    \end{tabular}
    }
\end{table}

\begin{table}[t]
    \centering
    \caption{Top 20 hashtags that \update{increased ($\beta > 0$)} in usage by QAnon post Twitter Action, estimated by regression discontinuity design in top 10K hashtags used by QAnon accounts} 
    \label{tab:qanon_rdd_emerging}
    \resizebox{0.48\textwidth}{!}{
    \begin{tabular}{l|c|c|c|c}
    \toprule
    \textbf{Hashtag} & $\bm{|\beta|}$ & slope & intercept & p-value \\
    \midrule 
walkawayfromdemocrats & 0.138 & -0.001 & 0.145 & \textbf{0.000} \\ 
saveourchildren & 0.099 & 0.002 & 0.011 & 0.158 \\ 
huge & 0.074 & -0.001 & 0.016 & 0.130 \\ 
heelsupharris & 0.068 & 0.0 & 0.006 &  0.098 \\ 
vote & 0.067 & 0.0 & 0.071 & \textbf{0.005} \\ 
warroompandemic & 0.065 & -0.001 & 0.041 & \textbf{0.007} \\ 
th3d3n & 0.064 & -0.001 & 0.016 & \textbf{0.002} \\ 
womenfortrump & 0.061 & 0.0 & 0.086 & 0.133 \\ 
thesilentmajority & 0.059 & 0.0 & 0.017 & 0.122 \\ 
sallyyates & 0.058 & -0.001 & 0.015 &   0.058 \\ 
hcqworksfauciknewin2005 & 0.055 & -0.001 & 0.014 & \textbf{0.000} \\ 
nomailinvoting & 0.053 & 0.0 & 0.014 &  \textbf{0.000} \\ 
mo03 & 0.051 & -0.001 & 0.015 & \textbf{0.032} \\ 
trump2020victory & 0.05 & -0.001 & 0.03 & \textbf{0.001} \\ 
hermancain & 0.05 & -0.001 & 0.016 &  \textbf{0.003} \\ 
bigpharma & 0.05 & -0.001 & 0.018 & \textbf{0.000} \\ 
jimcrowjoe & 0.05 & -0.001 & 0.032 &  0.072 \\ 
bidenisapedo & 0.049 & -0.001 & 0.014 &  \textbf{0.011} \\ 
vppick & 0.049 & -0.001 & 0.008 & 0.172 \\ 
hcqzinc4prevention & 0.048 & -0.001 & 0.015 & \textbf{0.000} \\ 
    \bottomrule
    \end{tabular}
    }
\end{table}

\begin{table}[t]
    \centering
    \caption{\% Decrease in ratio of direct engagements (reply to, retweet, quote) with QAnon accounts tweets to volume of QAnon accounts tweets, before and after Twitter Action}
    \label{tab:engchange}
    \resizebox{0.45\textwidth}{!}{
    \begin{tabular}{c|p{1.2cm}|p{1.2cm}|p{1.5cm}|p{2cm}}
    \toprule
    Tweet Type & Vol./ day (Before) &  Vol./ day (After) & \%Increase (Vol./ day) & \%Decrease (Eng vol./ QAnon tweet)
    \\
    \midrule
    QAnon & 295,267 & 361,176 & 22.32 & - \\
    RP$\_$TO & 49,161 & 46,114 & -6.2 & 23.32 \\
    RT & 177,013 & 189,387 & 6.99 & 12.53 \\
    QTD & 8908 & 8985 & 0.86 &  17.55 \\
    \bottomrule
    \end{tabular}
    }
\end{table}

Twitter restrictions attempted to make QAnon content less visible, by banning QAnon-related content from appearing in trends and search. Therefore, we examine changes in QAnon content through analysis of hashtag usage patterns, that could have been employed as evasion strategies to sidestep Twitter restrictions. To that end, we leverage \textit{regression discontinuity design} (RDD) to estimate causal effects of Twitter intervention, on hashtags adopted by QAnon accounts. RDD is a design which elicits causal effects of interventions by assigning a threshold above/below which an intervention is assigned \citep{lee2010regression}. By comparing observations lying closely on either side of the threshold, it is possible to estimate the average treatment effect from observed data, when randomized trials are infeasible.

For RDD, we consider each hashtag in the most frequent 10K hashtags adopted in QAnon account tweets, and fit a linear regression model on the hashtag's daily usage volume during the data collection period, with an additional treatment variable assigned to the regression model to capture intervention effect. The regression model is as follows, $y = mx + b + \beta \mathbb{I}\{x > x_0\}$. Here, the hashtag usage volume per day (i.e., how many times the hashtag was adopted in tweets from QAnon accounts), normalized by tweet volume from QAnon accounts on that day, is modeled by dependent variable $y$, over each day $x$. The slope $m$ and intercept $b$ captures the trend in hashtag usage, and the coefficient $\beta$ of the treatment variable $\mathbb{I}\{x > x_0\}$ captures the discontinuity at threshold $x_0$, selected as the end of the week that Twitter announced and enforced its restrictions on QAnon.

In Table~\ref{tab:qanon_rdd_declining} and \ref{tab:qanon_rdd_emerging}, we list identified hashtags with highest estimated treatment effects $|\beta|$ (i.e., most change with respect to Twitter's intervention on QAnon with p-value of $\beta$ at most 0.2) and highlighted in bold are p-value $\leq 0.05$ i.e., 95\% confidence interval.\footnote{\update{The goodness of fit tests and other regression statistics, with few illustrated examples of identified hashtags are in the Appendix.}} \update{In the assumed function, since a common slope with different intercepts ($b$ before intervention and $b + \beta$ after intervention) is used to capture the estimated treatment effect, both coefficients $\beta$ and $m$ determine whether the hashtag's usage declined or increased post intervention. For declining hashtags, $\beta < 0, m \leq 0$ are indicators of decreased intercept post intervention, whereas $\beta > 0, m \geq 0$ indicate increased intercept and increased hashtag usage post intervention. An error margin of $\pm0.001$ from $0$ was considered for the slope based on inspection of scatter plots of data for fitted regressions that were nearly parallel to the x-axis, resulting in difference of averages on either side of the intervention as estimated treatment effects.}

Declining hashtags post intervention (ordered by magnitude $|\beta|$) included prominent tags associated with the QAnon movement such as wwg1wga, qanon, q, qarmy, deepstate (Table~\ref{tab:qanon_rdd_declining}). Increasing hashtags post intervention (Table~\ref{tab:qanon_rdd_emerging}) included alternate hashtags e.g. walkawayfromdemocrats, nomailinvoting, bidenisapedo, hcqworksfauciknewin2005, suggesting a disconcerting gap in intended enforcement actions and their actual effectiveness.


We  examined the change in volume of engagements (direct engagements in the form of replies to, retweets of, quotes of) with QAnon accounts tweets, following Twitter intervention on QAnon (Tab~\ref{tab:engchange}). While the volume of retweet engagements per day per
QAnon account tweet decreased by 12.53\%, from earlier part of
the timeline to later half after Twitter action, the overall volume
of retweet engagements per day increased by $\sim$7\%. This can be
explained by 22.32\% increase in QAnon tweet volume per day,
compensating for decrease in engagements per tweet. Similar
for quoted tweets, but for reply tweets the engagements volume
declined overall by $\sim$6\% in the later half of the timeline. 

\begin{figure*}[t]
     \centering
     \begin{subfigure}[b]{0.23\textwidth}
         \centering
         \includegraphics[width=\textwidth,height=3.5cm]{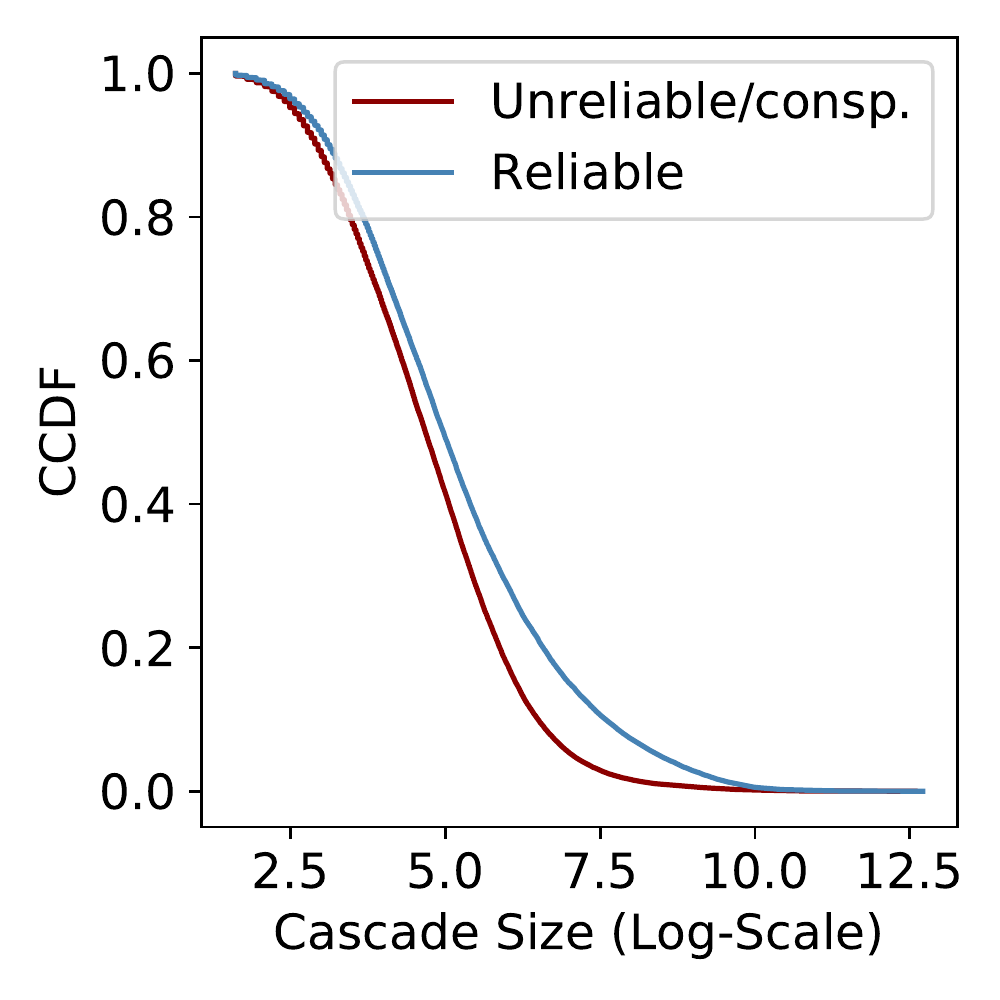}
        \caption{}
        \label{fig:ccdf}
     \end{subfigure}
     ~
     \begin{subfigure}[b]{0.23\textwidth}
         \centering
         \includegraphics[width=\textwidth,height=3.5cm]{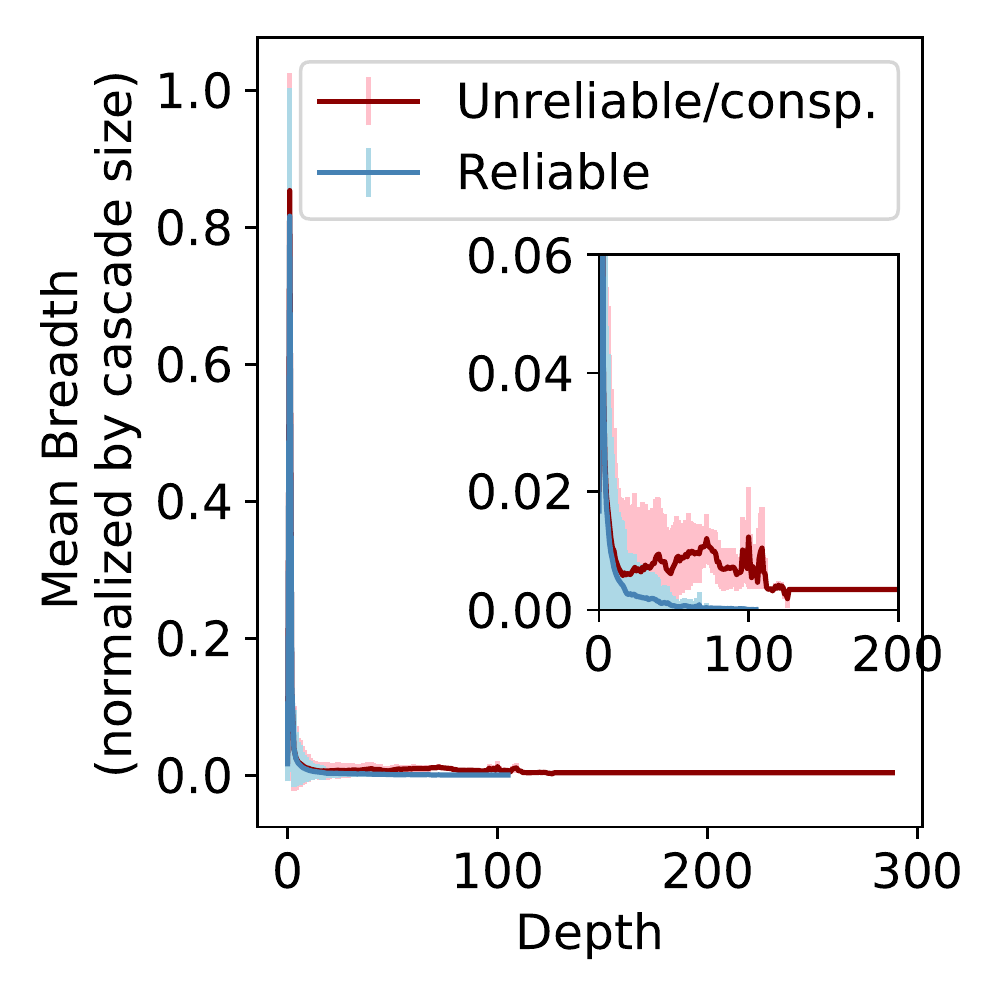}
         \caption{}
         \label{fig:breadth_depth}
     \end{subfigure}
    ~
     \begin{subfigure}[b]{0.23\textwidth}
         \centering
        \includegraphics[width=\textwidth,height=3.5cm]{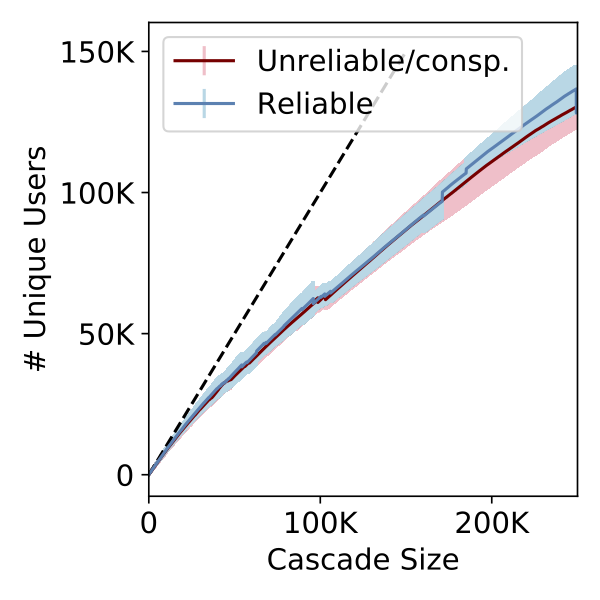}
        \caption{}
        \label{fig:uniq_users_cas_size}
     \end{subfigure}
     ~
     \begin{subfigure}[b]{0.23\textwidth}
         \centering
        \includegraphics[width=\textwidth,height=3.5cm]{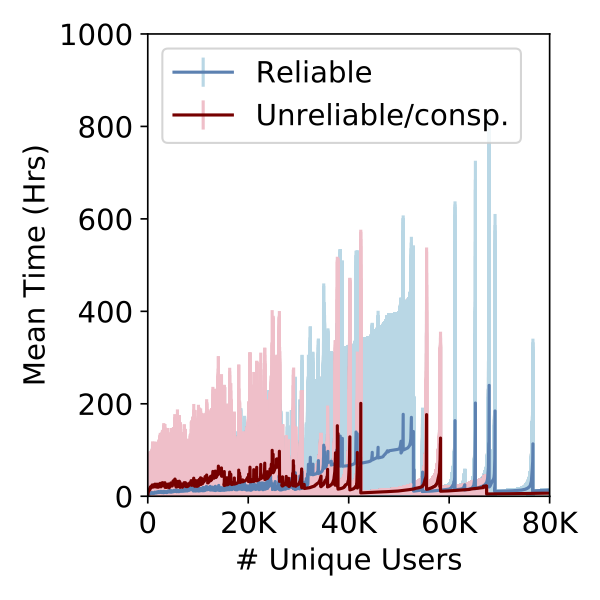}
         \caption{}
         \label{fig:mean_time_uniq_users}
     \end{subfigure}
    \caption{Comparison of information propagation dynamics of reliable vs. unreliable/conspiracy cascades identified in election-related tweets. (a) CCDF of cascade size. (b) Mean breadth to depth ratio. Reliable cascades run broader at shorter depths of cascade propagation trees. (c) Avg. unique users reached at each cascade size with more repeated engagements from same accounts in unreliable cascades. (d) Mean time to reach unique users is higher and more bursty for unreliable cascades.}
    \label{fig:cascade_props}
\end{figure*}

\subsection{Propagation dynamics of election cascades} 

In Fig.~\ref{fig:cascade_props}, we provide additional analysis of engagements with unreliable/conspiracy tweets by comparing cascade propagation dynamics. The engagements with the unreliable/conspiracy tweets appeared less viral (mean time to reach unique accounts is higher (Fig.~\ref{fig:mean_time_uniq_users}), and mean breadth of accounts reached at lesser depth of the propagation tree is smaller (Fig.~\ref{fig:breadth_depth})). \update{The propagation tree corresponding to each cascade was constructed from available retweet/reply/quote links between tweets.} Yet, there can be several unreliable/conspiracy cascades that do get many engagements (cascade size CCDF (Fig~\ref{fig:ccdf})). The findings about the propagation tree structure are similar to previous findings on unverified information or rumor cascades \citep{friggeri2014rumor}. In addition, unreliable/conspiracy cascades appear to have more repeated engagements (reaching fewer unique users for the same cascade size (Fig.~\ref{fig:uniq_users_cas_size})), which is also observed in \cite{ruchansky2017csi}. 

\update{\citeauthor{Vosoughi1146} (\citeyear{Vosoughi1146}) studied propagation dynamics of false and true news (verified on fact-checking websites e.g. snopes.com) on Twitter from 2006-2017. They found that fact-checked false news was more viral (diffused farther, faster, deeper and more broadly) than fact-checked true news. We note that the study provides useful findings, however, is specific to information that has been fact-checked (e.g. PolitiFact often fact-checks statements submitted by readers, and mentions that since they cannot check all claims, they select the most newsworthy and significant ones\footnote{\href{https://www.politifact.com/article/2018/feb/12/principles-truth-o-meter-politifacts-methodology-i/\#How\%20we\%20choose\%20claims}{PolitiFact.com/How we choose claims to fact-check}.}) Also, the fact-checked true rumors would likely not include mainstream news articles. In comparison, our findings about unreliable/conspiracy cascades being less viral than reliable cascades illustrate a more general comparison, beyond only popular false or non-mainstream true news.}

\section{Discussions and Conclusions}

In this work, we studied the disinformation landscape through identification of unreliable/conspiracy tweets, and analysis of QAnon conspiracy group and its activities. We focused on characterization of targeted topics of disinformation, and engagements with QAnon based on political leaning and tweet types, to understand how attempts to manipulate the discourse prior to the election were conducted. Unfortunately, our findings also suggest that Twitter actions to limit the QAnon conspiracy might not have been sufficient and effective, and even known conspiracy groups can adapt and evade imposed restrictions.

The characterization of engagements based on political leaning also indicates that many accounts in the active accounts (appearing more than 20 times in collected tweets) had interactions with QAnon account tweets, with majority of the endorsements being from right-leaning accounts. QAnon also actively engaged in discussions with left-leaning accounts through replies. Conspiratorial and far-right or far-left narratives can therefore potentially worsen the ideological separation in U.S. politics. Even certain unreliable and conspiracy tweets which contain misleading information and lack evidence, can get significant reshares, and increase uncertainty about the truth.

In conclusion, we discuss limitations of the current study. Although we find high recall of unreliable and conspiracy claims with the CSI model based on cross-validation over tweets labeled using news source URLs, reducing false positives is desirable for solutions to limit disinformation, without censoring free speech. Also, distorted narratives lie on a spectrum of truth, that can be hard to assess. In addition, here we regard right-leaning accounts that tweet pre-defined QAnon related keywords or hashtags as QAnon accounts. In future work, it will be desirable to make finer distinctions to characterize conspiratorial vs. non-conspiratorial far-right narratives in QAnon tweets, as well as to characterize tweets that debunk QAnon conspiracies, and investigate account interactions based on these distinctions. \update{We also discuss the limitations of causal estimation from observational data. In RDD, since we do not control for confounding variables that might be related to real-world events, it is possible that a spike caused in a hashtag's usage near the intervention might be attributed to the intervention, instead of a possible unrelated confounding event. Also, given the certainty of the Twitter action date, we use RDD with a hard threshold, which is a reasonable assumption based on inspection of the results; however, extensions to fuzzy RDD  could be alternatively considered \cite{lee2010regression}.}

\update{We also discuss inherent limitations in data collection. The standard API allows access to a $\sim$1\% sample of the Twitter stream filtered based on tracked parameters.
Since Twitter does not allow unpaid access to historical tweets older than 7 days, it is hard to do iterative/snowball sampling based  on  keyword  expansions.   Therefore,  we  rely  on tracking candidate mentions in tweet text/metadata instead of manually selected fixed keywords. The inspected topics, accounts, and hashtags distributions do suggest comprehensive coverage of election events and disinformation.}

\section{Appendix}



\begin{figure}
    \centering
    \includegraphics[width=0.48\textwidth]{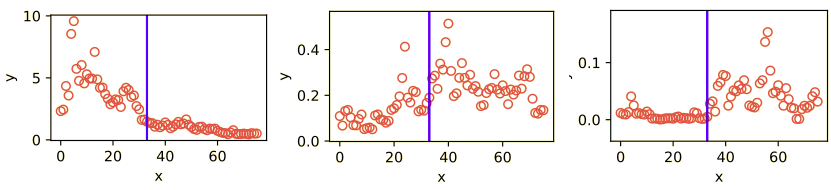}
    \caption{RDD data plot for example hashtags (Left) ``wwg1wga" (Middle) ``walkawayfromdemocrats" (Right) ``nomailinvoting" before and after Twitter action (fraction of hashtag usage in volume of QAnon tweets 'y' vs. day 'x')}
    \label{fig:illustrate_rdd}
\end{figure}

\begin{table}[t]
    \centering
    \caption{\update{Regression stats. and goodness of fit for different degree polynomials. Note: *(p-value $<=$ 0.05). AIC, BIC were lower for degree-1 compared to higher degree curves.}}
    \label{tab:regr_rdd_stats}
    \resizebox{0.49\textwidth}{!}{
    \begin{tabular}{l|c|c|c|c|c|c|c|c}
    \toprule
    Hashtag & $R^2$ & $R^2$adj & F-stat & AIC & BIC & $\beta$ & m & $m_2$ \\
    \midrule
wwg1wga & 0.72 & 0.71 & 93.64$^*$ & 229.71 & 236.7 & -1.51$^*$ & -0.05$^*$ &  \\
 & 0.72 & 0.71 & 62.87$^*$ & 230.57 & 239.89 & -1.5$^*$ & -0.11 & 0.06 \\
qanon & 0.68 & 0.67 & 78.22$^*$ & 172.67 & 179.66 & -0.90$^*$ & -0.03$^*$ &  \\
 & 0.68 & 0.67 & 51.81$^*$ & 174.29 & 183.61 & -0.89$^*$ & -0.06 & 0.03 \\
kag & 0.51 & 0.49 & 37.37$^*$ & 1.68 & 8.67 & -0.4q$^*$ & -0.0 &  \\
 & 0.51 & 0.49 & 24.6$^*$ & 3.64 & 12.96 & -0.4$^*$ & -0.0 & 0.0 \\
q & 0.67 & 0.67 & 75.77$^*$ & -90.17 & -83.17 & -0.23$^*$ & -0.0$^*$ &  \\
 & 0.68 & 0.66 & 50.24$^*$ & -88.6 & -79.27 & -0.22$^*$ & -0.01 & 0.0 \\
qarmy & 0.58 & 0.57 & 50.28$^*$ & -4.88 & 2.11 & -0.21$^*$ & -0.01$^*$ &  \\
 & 0.58 & 0.56 & 33.16$^*$ & -3.01 & 6.31 & -0.21 & -0.01 & 0.0 \\
qanons & 0.58 & 0.56 & 49.57$^*$ & -119.35 & -112.36 & -0.11$^*$ & -0.0$^*$ &  \\
 & 0.58 & 0.57 & 33.73$^*$ & -118.86 & -109.54 & -0.11$^*$ & -0.01 & 0.01 \\
patriotstriketeam & 0.45 & 0.44 & 30.39$^*$ & -183.19 & -176.2 & -0.09$^*$ & -0.0 &  \\
 & 0.45 & 0.43 & 19.98$^*$ & -181.19 & -171.87 & -0.08$^*$ & -0.0 & 0.0 \\
deepstate & 0.76 & 0.75 & 113.22$^*$ & -284.36 & -277.36 & -0.08$^*$ & -0.0$^*$ &  \\
 & 0.76 & 0.75 & 74.52$^*$ & -282.41 & -273.08 & -0.08$^*$ & -0.0 & -0.0 \\
walkawayfromdemocrats & 0.36 & 0.34 & 20.67$^*$ & -174.36 & -167.37 & 0.14$^*$ & -0.0 &  \\
 & 0.36 & 0.34 & 13.71$^*$ & -172.61 & -163.28 & 0.14$^*$ & -0.0 & 0.0 \\
vote & 0.19 & 0.17 & 8.79$^*$ & -229.96 & -222.96 & 0.07$^*$ & -0.0 &  \\
 & 0.2 & 0.17 & 6.07$^*$ & -228.7 & -219.38 & 0.07$^*$ & -0.0 & 0.0 \\
warroompandemic & 0.11 & 0.08 & 4.3$^*$ & -229.15 & -222.15 & 0.06$^*$ & -0.0 &  \\
 & 0.11 & 0.07 & 2.84$^*$ & -227.19 & -217.86 & 0.07$^*$ & -0.0 & -0.0 \\
th3d3n & 0.15 & 0.13 & 6.41$^*$ & -259.06 & -252.07 & 0.06$^*$ & -0.0$^*$ &  \\
 & 0.18 & 0.14 & 5.22$^*$ & -259.73 & -250.41 & 0.06$^*$ & -0.0 & 0.0 \\
hcqworksfauciknewin2005 & 0.38 & 0.36 & 22.45$^*$ & -380.48 & -373.49 & 0.06$^*$ & -0.0$^*$ &  \\
 & 0.38 & 0.36 & 15.0$^*$ & -378.94 & -369.62 & 0.06$^*$ & -0.0 & 0.0 \\
nomailinvoting & 0.41 & 0.4 & 25.48$^*$ & -351.56 & -344.57 & 0.05$^*$ & -0.0 &  \\
 & 0.41 & 0.39 & 16.78$^*$ & -349.6 & -340.28 & 0.05$^*$ & -0.0 & 0.0 \\
mo03 & 0.06 & 0.04 & 2.42 & -230.76 & -223.77 & 0.05$^*$ & -0.0 &  \\
 & 0.09 & 0.05 & 2.28 & -230.79 & -221.46 & 0.05$^*$ & -0.01 & 0.0 \\
trump2020victory & 0.18 & 0.15 & 7.78$^*$ & -303.01 & -296.01 & 0.05$^*$ & -0.0 &  \\
 & 0.18 & 0.14 & 5.23$^*$ & -301.3 & -291.98 & 0.05$^*$ & -0.0 & 0.0 \\
hermancain & 0.12 & 0.09 & 4.91$^*$ & -270.0 & -263.01 & 0.06$^*$ & -0.0$^*$ &  \\
 & 0.12 & 0.09 & 3.42$^*$ & -268.52 & -259.2 & 0.06$^*$ & 0.0 & -0.0 \\
bigpharma & 0.31 & 0.29 & 16.61$^*$ & -371.12 & -364.12 & 0.05$^*$ & -0.0$^*$ &  \\
 & 0.31 & 0.28 & 10.94$^*$ & -369.17 & -359.85 & 0.05$^*$ & -0.0 & -0.0 \\
bidenisapedo & 0.09 & 0.07 & 3.71$^*$ & -265.18 & -258.19 & 0.05$^*$ & -0.0 &  \\
 & 0.11 & 0.08 & 3.06$^*$ & -264.93 & -255.61 & 0.05$^*$ & -0.0 & 0.0 \\
hcqzinc4prevention & 0.24 & 0.22 & 11.58$^*$ & -358.27 & -351.28 & 0.05$^*$ & -0.0$^*$ &  \\
 & 0.24 & 0.21 & 7.62$^*$ & -356.27 & -346.95 & 0.05$^*$ & -0.0 & 0.0 \\
    \bottomrule
    \end{tabular}
    }
\end{table}

\textbf{Regression statistics.}
\update{In Table~\ref{tab:regr_rdd_stats}, goodness-of-fit tests are provided for the RDD regression function (illustrated examples in Fig~\ref{fig:illustrate_rdd}). When comparing different degree polynomials as chosen functions for RDD, we found AIC and BIC were lowest for degree-1 (we measured up to degree-4, since the results are similar we show only degree-1 and degree-2 in the table). Degree-1 is the model used in the RDD analysis $y = mx + b + \beta \mathcal{I}(x > x_0)$ \cite{lee2010regression}. Adjusted $R^2$, f-statistic and p-values of the coefficients also indicate degree-1 is preferred, since the p-value of coefficient $m_2$ for $x^2$ term is not significant at 0.05. The results were also consistent when separate coefficients of dependent variables were used on either side of the intervention.}


\begin{figure}[t]
    \centering
    \includegraphics[width=0.5\textwidth]{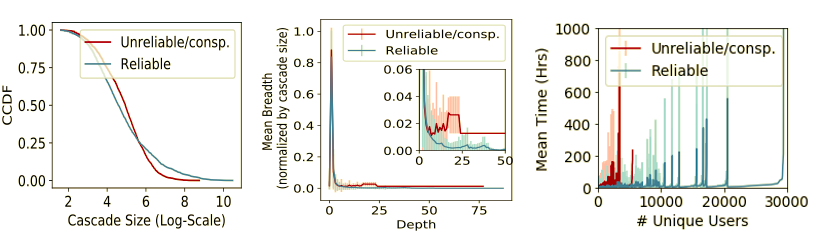}
    \caption{Cascade propagation comparison only on labeled set (unreliable/conspiracy vs. reliable news source labels).}
    \label{fig:extra_cascade_props}
\end{figure}

\noindent \textbf{Cascade propagation.} \update{
To confirm that our findings are not biased by CSI predictions, in Fig~\ref{fig:extra_cascade_props}, we compare only labeled set of unreliable/conspiracy vs. reliable cascades (labeled by the news source, used in training CSI). We find the conclusions are robust with similar trends for all propagation properties. CSI uses engagements, but also text, temporal and account suspiciousness features, making predictions more robust, and verified to be within reasonable error rates.}


\fontsize{9pt}{8pt} \selectfont
\bibliography{references}

\begin{thebibliography}{31}
\providecommand{\natexlab}[1]{#1}
\providecommand{\url}[1]{\texttt{#1}}
\providecommand{\urlprefix}{URL }
\expandafter\ifx\csname urlstyle\endcsname\relax
  \providecommand{\doi}[1]{doi:\discretionary{}{}{}#1}\else
  \providecommand{\doi}{doi:\discretionary{}{}{}\begingroup
  \urlstyle{rm}\Url}\fi

\bibitem[{Badawy et~al.(2019)Badawy, Addawood, Lerman, and
  Ferrara}]{badawy2019characterizing}
Badawy, A.; Addawood, A.; Lerman, K.; and Ferrara, E. 2019.
\newblock Characterizing the 2016 Russian IRA influence campaign.
\newblock \emph{SNAM} .

\bibitem[{Bessi and Ferrara(2016)}]{bessi2016social}
Bessi, A.; and Ferrara, E. 2016.
\newblock Social bots distort the 2016 US Presidential election online
  discussion.
\newblock \emph{First Monday} 21(11).

\bibitem[{Blondel et~al.(2008)Blondel, Guillaume, Lambiotte, and
  Lefebvre}]{blondel2008fast}
Blondel, V.~D.; Guillaume, J.-L.; Lambiotte, R.; and Lefebvre, E. 2008.
\newblock Fast unfolding of communities in large networks.
\newblock \emph{Journal of statistical mechanics: theory and experiment}
  2008(10): P10008.

\bibitem[{Bojanowski et~al.(2017)Bojanowski, Grave, Joulin, and
  Mikolov}]{bojanowski-etal-2017-enriching}
Bojanowski, P.; Grave, E.; Joulin, A.; and Mikolov, T. 2017.
\newblock Enriching Word Vectors with Subword Information.
\newblock \emph{TACL} 5.

\bibitem[{Chen, Deb, and Ferrara(2020)}]{chen2020election2020}
Chen, E.; Deb, A.; and Ferrara, E. 2020.
\newblock \#Election2020: The First Public Twitter Dataset on the 2020 US
  Presidential Election.

\bibitem[{Cresci(2020)}]{Cresci_2020}
Cresci, S. 2020.
\newblock A decade of social bot detection.
\newblock \emph{ACM} 63(10): 72–83.
\newblock ISSN 1557-7317.

\bibitem[{de~Zeeuw et~al.(2020)de~Zeeuw, Hagen, Peeters, and
  Jokubauskaite}]{de2020tracing}
de~Zeeuw, D.; Hagen, S.; Peeters, S.; and Jokubauskaite, E. 2020.
\newblock Tracing normiefication.
\newblock \emph{First Monday} .

\bibitem[{Ferrara et~al.(2020)Ferrara, Chang, Chen, Muric, and
  Patel}]{ferrara2020characterizing}
Ferrara, E.; Chang, H.; Chen, E.; Muric, G.; and Patel, J. 2020.
\newblock Characterizing social media manipulation in the 2020 US presidential
  election.
\newblock \emph{First Monday} .

\bibitem[{Friggeri et~al.(2014)Friggeri, Adamic, Eckles, and
  Cheng}]{friggeri2014rumor}
Friggeri, A.; Adamic, L.~A.; Eckles, D.; and Cheng, J. 2014.
\newblock Rumor Cascades.
\newblock In \emph{ICWSM}.

\bibitem[{Gadde and Roth(2018)}]{gadde2018enabling}
Gadde, V.; and Roth, Y. 2018.
\newblock Enabling further research of information operations on Twitter.
\newblock \emph{Twitter Blog} 17.

\bibitem[{Hughes and Wojcik(2019 (accessed March 20, 2020))}]{PewTwitter2019}
Hughes, A.; and Wojcik, S. 2019 (accessed March 20, 2020).
\newblock \emph{10 facts about Americans and Twitter}.
\newblock
  \urlprefix\url{https://www.pewresearch.org/fact-tank/2019/08/02/10-facts-about-americans-and-twitter/}.

\bibitem[{Jolley, Meleady, and Douglas(2020)}]{jolley2020exposure}
Jolley, D.; Meleady, R.; and Douglas, K.~M. 2020.
\newblock Exposure to intergroup conspiracy theories promotes prejudice which
  spreads across groups.
\newblock \emph{British Journal of Psychology} 111(1): 17--35.

\bibitem[{Lee and Lemieux(2010)}]{lee2010regression}
Lee, D.~S.; and Lemieux, T. 2010.
\newblock Regression discontinuity designs in economics.
\newblock \emph{Journal of economic literature} 48(2): 281--355.

\bibitem[{Li et~al.(2016)Li, Wang, Zhang, Sun, and Ma}]{li2016topic}
Li, C.; Wang, H.; Zhang, Z.; Sun, A.; and Ma, Z. 2016.
\newblock Topic modeling for short texts with auxiliary word embeddings.
\newblock In \emph{SIGIR}.

\bibitem[{Loader and Mercea(2011)}]{loader2011networking}
Loader, B.~D.; and Mercea, D. 2011.
\newblock Networking democracy? Social media innovations and participatory
  politics.
\newblock \emph{ICS} 14(6).

\bibitem[{Luceri, Cardoso, and Giordano(2020)}]{luceri2020down}
Luceri, L.; Cardoso, F.; and Giordano, S. 2020.
\newblock Down the bot hole: actionable insights from a 1-year analysis of bots
  activity on Twitter.
\newblock \emph{arXiv preprint arXiv:2010.15820} .

\bibitem[{Ma et~al.(2016)Ma, Gao, Mitra, Kwon, Jansen, Wong, and
  Cha}]{ma2016detecting}
Ma, J.; Gao, W.; Mitra, P.; Kwon, S.; Jansen, B.~J.; Wong, K.-F.; and Cha, M.
  2016.
\newblock Detecting Rumors from Microblogs with Recurrent Neural Networks.
\newblock In \emph{IJCAI}, 3818--3824.

\bibitem[{Martin and Shapiro(2019)}]{martin2019trends}
Martin, D.~A.; and Shapiro, J.~N. 2019.
\newblock Trends in online foreign influence efforts.

\bibitem[{Miles~P.(2020 (accessed June 18, 2020))}]{NPRForeign}
Miles~P., P.~E. 2020 (accessed June 18, 2020).
\newblock \emph{Foreign Interference Persists And Techniques Are Evolving, Big
  Tech Tells Hill}.
\newblock
  \urlprefix\url{https://www.npr.org/2020/06/18/880349422/foreign-interference-persists-and-techniques-are-evolving-big-tech-tells-hill}.

\bibitem[{Nyhan and Reifler(2010)}]{nyhan2010corrections}
Nyhan, B.; and Reifler, J. 2010.
\newblock When corrections fail: The persistence of political misperceptions.
\newblock \emph{Political Behavior} 32(2).

\bibitem[{Papasavva et~al.(2020)Papasavva, Blackburn, Stringhini, Zannettou,
  and De~Cristofaro}]{papasavva2020qoincidence}
Papasavva, A.; Blackburn, J.; Stringhini, G.; Zannettou, S.; and De~Cristofaro,
  E. 2020.
\newblock " Is it a Qoincidence?": A First Step Towards Understanding and
  Characterizing the QAnon Movement on Voat. co.
\newblock \emph{arXiv preprint arXiv:2009.04885} .

\bibitem[{Phadke, Samory, and Mitra(2021)}]{phadke2021makes}
Phadke, S.; Samory, M.; and Mitra, T. 2021.
\newblock What Makes People Join Conspiracy Communities? Role of Social Factors
  in Conspiracy Engagement.
\newblock \emph{ACM HCI} 4(CSCW3): 1--30.

\bibitem[{Ruchansky, Seo, and Liu(2017)}]{ruchansky2017csi}
Ruchansky, N.; Seo, S.; and Liu, Y. 2017.
\newblock CSI: A Hybrid Deep Model for Fake News Detection.
\newblock In \emph{CIKM}, 797--806. ACM.

\bibitem[{Sharma, Ferrara, and Liu(2020)}]{sharma2020identifying}
Sharma, K.; Ferrara, E.; and Liu, Y. 2020.
\newblock Identifying Coordinated Accounts in Disinformation Campaigns.
\newblock \emph{arXiv preprint arXiv:2008.11308} .

\bibitem[{Sharma et~al.(2019)Sharma, Qian, Jiang, Ruchansky, Zhang, and
  Liu}]{sharma2019combating}
Sharma, K.; Qian, F.; Jiang, H.; Ruchansky, N.; Zhang, M.; and Liu, Y. 2019.
\newblock Combating Fake News: A Survey on Identification and Mitigation
  Techniques.
\newblock \emph{ACM TIST} .

\bibitem[{Silva et~al.(2020)Silva, Ceschin, Shrestha, Brant, Fernandes, Silva,
  Gr{\'e}gio, Oliveira, and Giovanini}]{silva2020predicting}
Silva, M.; Ceschin, F.; Shrestha, P.; Brant, C.; Fernandes, J.; Silva, C.~S.;
  Gr{\'e}gio, A.; Oliveira, D.; and Giovanini, L. 2020.
\newblock Predicting Misinformation and Engagement in COVID-19 Twitter
  Discourse in the First Months of the Outbreak.
\newblock \emph{arXiv preprint arXiv:2012.02164} .

\bibitem[{Torres-Lugo, Yang, and Menczer(2020)}]{torreslugo2020manufacture}
Torres-Lugo, C.; Yang, K.-C.; and Menczer, F. 2020.
\newblock The Manufacture of Political Echo Chambers by Follow Train Abuse on
  Twitter.

\bibitem[{Van~der Linden(2015)}]{van2015conspiracy}
Van~der Linden, S. 2015.
\newblock The conspiracy-effect: Exposure to conspiracy theories (about global
  warming) decreases pro-social behavior and science acceptance.
\newblock \emph{Personality and Individual Differences} 87: 171--173.

\bibitem[{Vosoughi, Roy, and Aral(2018)}]{Vosoughi1146}
Vosoughi, S.; Roy, D.; and Aral, S. 2018.
\newblock The spread of true and false news online.
\newblock \emph{Science} 359(6380): 1146--1151.
\newblock ISSN 0036-8075.
\newblock \doi{10.1126/science.aap9559}.
\newblock \urlprefix\url{https://science.sciencemag.org/content/359/6380/1146}.

\bibitem[{Woolley and Howard(2017)}]{woolley2017computational}
Woolley, S.~C.; and Howard, P. 2017.
\newblock Computational propaganda worldwide: Executive summary.
\newblock \emph{Oxford Internet Institute} .

\bibitem[{Zimdars(2016)}]{zimdars2016false}
Zimdars, M. 2016.
\newblock \emph{False, Misleading, Clickbait-Y, and Satirical `News' Sources}.
\newblock
  \urlprefix\url{https://21stcenturywire.com/wp-content/uploads/2017/02/2017-DR-ZIMDARS-False-Misleading-Clickbait-y-and-Satirical-%E2%80%9CNews%E2%80%9D-Sources-Google-Docs.pdf}.

\end{thebibliography}

\end{document}